# Investigation of Mixing Characteristics in Strut Injectors using Modal Decomposition

Rahul Kumar Soni, [1)] Ashoke De[2, a)]

[1,2]Department of Aerospace Engineering, Indian Institute of Technology Kanpur, Kanpur, 208016, India

Effect of large-scale vortical structure on mixing and spreading of shear layer is numerically investigated. Two strut configurations namely Straight & Tapered strut at two convective Mach numbers ($M_c$ = 1.4 & 0.37) for two jet heights (0.6 & 1mm) are investigated. Hydrogen jet is injected through a two-dimensional slot in oncoming coflow at Mach 2. Excellent agreement between simulated and experimental data is witnessed, whereas the instantaneous data reveal the presence of various large-scale structures in the flow field. From the instantaneous field, it becomes apparent that both the geometries have different vortical breakdown locations. It is also noticed that an early onset of vortex breakdown manifests itself into the mixing layer thickness enhancement, effect of which is reflected in overall mixing characteristics. It becomes evident that the shear strength plays an important role in the near field mixing. The higher shear strength promotes the generation of large vortices. The analysis shows that SS-0.6 case offers highest mixing efficiency being dominated by relatively large-scale structures. Eigenmodes obtained through Proper Orthogonal Decomposition (POD) confirm the presence of dominating structures and shed light into the series of events involved in vortex pairing/merging and breakdown. Dynamic modal decomposition (DMD) also strengthens the observation made through the POD.

**I. INTRODUCTION**

The very notion of atmospheric flight at hypervelocity regime may appear fascinating and appealing but there are certain intrinsic challenges that need to be addressed. The concept of Scramjet was conceived as an air-breathing propulsion system, which is capable of propelling in high Mach number regime. But before the concept is realized certain questions need to be answered related to the design and technological issues. Although the proposed design may appear simple but achieving a stable and sustainable combustion efficiently at these speeds is an engineering marvel. The recent interest in Scramjet has promoted research towards the understanding of supersonic mixing phenomenon. The residence time being exceptionally short and need for efficient combustor has motivated research community to explore the underlying physics governing the mixing in these flow regimes.

Various injector configurations have been explored and studied, most popular being normal, oblique and parallel each of them is having their own pros and cons, while the detailed discussion can be found in [1-3]. Due to the simplicity of the

___

[a)] Author to whom correspondence should be addressed: ashoke@iitk.ac.in

design and contribution towards thrust, much effort is being put into the investigation of parallel injectors. The possibility of injecting the fuel into the core without inducing stronger shock wave as in normal injectors makes it an ideal candidate[4-6]. But it is also known that at the supersonic speeds the mixing capabilities are significantly reduced. This explains the need for research with various injectors especially simplistic in design yet highly efficient.

From the literature, it is well documented that majority of the studies have been performed using splitter plate. Brown & Roshko[7] investigated planar shear layer of two dissimilar gases for different velocity ratio ($R = u_2/u_1$) and density ratio ($S = \rho_2/\rho_1$), where they noted little influence of $S$ on mixing efficiency but demonstrated the presence of large-scale structures in the mixing layer. They also pointed out the reduction in spreading rate of the mixing layer with the increasing Mach number when compared to the spreading rate in the incompressible regime. This is due to the compressibility effects which inhibit the mass entrainment and turbulent diffusion. However, Gutmark et al.[8] in their investigation of rectangular and circular jets noticed an increase in spreading rate with increasing shear ($R$) irrespective of the convective Mach number ($M_c$). They also pointed out that these large-scale structures propagate at the convective Mach number defined in a frame of reference moving with the flow.

In the past, various studies [5, 9, 10, and 11] reported that the wake, generated due to the amplification of vortical structures, is present in the separated boundary layer and the base flow dominates the mixing process. It has been known for quite some time that the shock waves can be utilized to enhance the mixing. Because the shock wave (oblique) interacting with the jet of different density (relative to external flow) induces vorticity due to the non-alignment of pressure and density gradients. The production of vorticity enhances the strain rate and offers increased mixing due to diffusion. In this direction, the study conducted by Budzinsky et al.[12] investigated the effect of normal shock interaction with helium jet. They noticed increased molecular mixing rate by a factor of 2 for single interaction and around 3 times increment for double interactions. These findings strongly suggest that the multiple shock interactions must offer significant enhancement in mixing efficiency.

Furthermore, it is also established that the effects of compressibility manifest itself into the mixing efficiency through the convective Mach number. However, this manifestation due to convective Mach number ($M_c$) is also highly dependent on the injection technique deployed. It is argued that certain class of injectors relies on the two-dimensional wave growths for mixing enhancement, especially in low $M_c$ regime, whereas others dominated by oblique waves (three-dimensional) are effective at high $M_c$[3]. Therefore, it is quite imperative that the devices relying on production of axial vorticity such as ramp and strut type injectors are more effective at high $M_c$. The study by Yu et al.[13, 14] on cavity further bolstered the argument that the induced two-dimensional vortices were found to be more effective at low $M_c$, whereas the ramp injector, governed by three-dimensional vortices, exhibited mixing augmentation at higher $M_c$. Essentially the key observation from their study was that the two-dimensional vortices are more dominant in subsonic $M_c$, while the three-dimensional vortices play an important role for supersonic $M_c$ in mixing phenomenon.

From the above discussion, it is understood that the large-scale structures play an important role in the mixing escalation, especially where vorticity plays an important role. In turn, this suggests that the understanding of the coherent structures is vital to the overall understanding of any turbulent flow occurring in many engineering problems involving noise, mixing, drag etc. because they contain a large portion of the fluctuating energy. It is now widely accepted that the organized motions characterize the turbulent flows well. These organized structures or coherence has essential flow physics embedded into it. Therefore it becomes important to explore the behavior of these structures which would shed light into the turbulent flows and its control. The proper orthogonal decomposition (POD) is one such class of mathematical techniques that exploit the spatial orthogonality of the flow and reveals the spatial coherence. The Proper Orthogonal Decomposition (POD) was independently first proposed by Kosambi[15] and later extended to the study of turbulent flows by Lumley[16]. Later Sirovich[17] proposed computationally inexpensive "Method of Snapshot" which is applied in the present investigation to extract the dominating large-scale structures governing the mixing phenomenon. Previous works in the group[18, 19, 20, 21 and 22] extensively utilized this particular method in their studies related to backward facing step and base flow in the supersonic regime as well as in incompressible regime over the periodic hill and reported the presence of various large-scale structures in the shear layer and downstream regions. Although POD is a powerful tool to extract structures based on the fluctuating energy content but it suffers the loss of phase information, this means that the dynamic information embedded in the system is lost. This should suggest that even a small perturbation that can lead to instability will not be resolved in POD Eigen modes. Schmid[23] introduced another technique capable of extracting dynamic information namely Dynamic Mode Decomposition (DMD) based on Koopman analysis[24].

The major difference between POD and DMD lies in orthogonalization in space and time respectively. In DMD instead of fluctuating energy each mode is arranged based on the observed frequencies. So in this manner, the Eigen values of DMD represent growth or decay and oscillation frequencies of each mode. Therefore, in the present study dynamic decomposition is also performed to extract the dynamics of shear layer and wake mode in order to characterize the mixing phenomenon. Moreover, we have used wall modeled large eddy simulation methodology based on dynamic sgs modeling for two strut configurations namely Straight Strut (*SS*) and Tapered Strut (*TS*). The study is performed for two jet thickness 0.6 and 1 mm for jet Mach number ($M_j$) 2.3 and with 0.6 mm jet height for jet Mach number ($M_j$) 1. In the light of above survey, we attempt to understand the behavior of shear layer and its effect on the mixing. The present study also attempts to establish various parameters that affect mixing characteristics positively other than widely accepted convective Mach number. Best of the author's knowledge regarding literature, the modal decomposition techniques have never been used (do not exist) for this type of study and it is expected that these methodologies would shed more light into the underlying physics. The existence of large-scale structures coherent in space will be studied through the POD and DMD will shed light into the dynamics of the shear layer and instabilities present leading to the wake mode, which in turn affects the mixing characteristics.

## II. Numerical Details

### A. Governing Equations

The Favre-filtered governing equations for the conservation of mass, momentum, energy, and species are solved, as given by:

Continuity equation:

$$\frac{\partial}{\partial t}(\bar{\rho}) + \frac{\partial}{\partial x_i}(\bar{\rho}\tilde{u}_i) = 0 \tag{1}$$

Momentum equation:

$$\frac{\partial}{\partial t}(\bar{\rho}\tilde{u}_i) + \frac{\partial}{\partial x_j}(\bar{\rho}\tilde{u}_i\tilde{u}_j) = -\frac{\partial}{\partial x_i}(\bar{p}) + \frac{\partial}{\partial x_j}\left[(\mu + \mu_t)\frac{\partial \tilde{u}_i}{\partial x_j}\right] \tag{2}$$

Energy equation:

$$\frac{\partial}{\partial t}(\bar{\rho}\tilde{E}) + \frac{\partial}{\partial x_i}(\bar{\rho}\tilde{u}_i\tilde{E}) = -\frac{\partial}{\partial x_j}\left[\tilde{u}_j\left(-\tilde{p}I + \mu\frac{\partial \tilde{u}_i}{\partial x_j}\right)\right] + \frac{\partial}{\partial x_i}\left[\left(k + \frac{\mu_t C_p}{\text{Pr}_t}\right)\frac{\partial \tilde{T}}{\partial x_i}\right] \tag{3}$$

Species Transport:

$$\frac{\partial}{\partial t}\left(\bar{\rho}\tilde{Y}_k\right) + \frac{\partial}{\partial x_j}\left(\bar{\rho}\tilde{u}_j\tilde{Y}_k\right) - \frac{\partial}{\partial x_j}\left[\bar{\rho}\left(D_k + \frac{\nu_t}{Sc_t}\right)\frac{\partial \tilde{Y}_k}{\partial x_j}\right] = 0 \tag{4}$$

Where (~) and (-) in above equation refer to filtered and Favre filtered quantity, $\rho$ is the density, $u_i$ is the velocity vector, p is the pressure, $E = e + u_i^2/2$ is the total energy, where $e = h - p/\rho$ is the internal energy and $h$ is enthalpy. The fluid properties $\mu$, $D$ and $k$ are respectively the viscosity, mass diffusivity and the thermal conductivity, while $\mu_t$, $Sc_t$ and $P_{rt}$ are the turbulent eddy viscosity, turbulent Schmidt number and turbulent Prandtl number respectively. The turbulent Prandtl and Schmidt numbers are defined as 0.9 & 0.5 and the dynamic viscosity is evaluated through the Sutherland's law,

$$\mu = \frac{A_s \sqrt{T}}{1 + \frac{T_s}{T}} \tag{5}$$

Here $A_s$ is the Sutherland coefficient and $T_s$ is Sutherland temperature values of which are provided in Table I.

Table I. Sutherland coefficients and Temperature for species utilized in computation

| Species | $A_s$ | $T_s$ [K] |
|---------|-----------|-----------|
| $H_2$   | 8.411e-06 | 96        |
| $N_2$   | 1.458e-06 | 110.4     |
| $O_2$   | 1.458e-06 | 110.4     |

The specific heat at constant pressure $c_p$ is evaluated through the following polynomial

$$c_p = R\left(\left(\left(\left(a_4 T + a_3\right)T + a_2\right)T + a_1 T\right) + a_0\right) \qquad (6)$$

Additionally, $a_5$ & $a_6$ are the constants of integration at low and high temperature end for evaluation of enthalpy and entropy. The numerical values of the coefficients ($a_i$) can be accessed from the work of Goos et al.[25]. Finally the above sets of governing equations are closed by including the ideal gas equation.

The turbulence is modeled through the Large Eddy Simulation (LES) approach, where the small scale motions are modeled while the large-scale motions are resolved by filtering the instantaneous governing equations. In the present study, wall modeled dynamic LES model based on two model coefficients is invoked where the gradient approximation is used to relate the unresolved stresses to re-solved velocity field and given as: $\tau_{ij}^{sgs} - \frac{1}{3}\delta_{ij}\tau_{kk} = -2\bar{\rho}C\bar{\Delta}^2(\tilde{S}_{ij} - \frac{1}{3}\delta_{ij}\tilde{S}_{ij})$ where $\tau_{kk}^{sgs} = 2\bar{\rho}C_I\bar{\Delta}^2\left|\tilde{S}_{ij}\right|$, here ~ and − refer to filtered and Favre averaged variables, $S_{ij}$ represents the strain rate tensor. The $C$ & $C_I$ are the model coefficients are evaluated dynamically and averaged locally. Further details about the model and wall modelling approach used herein can be found in the work of Soni et al.[18].

**B. Proper Orthogonal Decomposition**

The proper orthogonal decomposition technique involves modal decomposition based on optimization of the field variables under examination. It is well known fact that turbulence has infinite degree of freedom due to the presence of the infinite scales and analyzing such system is not possible. This is where POD comes into the picture; modal decomposition reveals the order and structures (or pattern) in the turbulent flow field. This reduces the system to lower dimension based on the energy distribution across the modes. So, this way only first few $N$ (based on first $N$ eigenvalues) modes possessing higher energy are needed to characterize the turbulence. Mathematically, POD attempts to search for set of orthonormal basis vectors in a lower dimension by decomposing ensemble of realizations while minimizing the error between the original data and its projection onto the lower dimensional space. The mathematical description of this decomposition technique can be found in literature[18-20] as briefly described here.

Let $U(x,t)$ represents a fluctuating vector field with three components while considering $N$ snapshots in time. Assuming the ensemble is sufficiently large, the auto correlation tensor $R(x,x')$ can be approximated as the following

$$R(x,x') = \frac{1}{N}\sum_{n=1}^{N} U(x,t^n)U^T(x',t^n) \tag{7}$$

Let us assume the basis mode can be written in terms of original data set as given below,

$$\emptyset(x) = \sum_{n=1}^{N} A(t^n)U(x,t^n) \tag{8}$$

Using $R(x,x')$ & $\emptyset(x)$, the eigen value problem can be written as

$$\sum_{n=1}^{N}\left(\frac{1}{N}\int_{\Omega} U^T(x',t^n)U(x',t^n)dx'\right)A(t^n) = \lambda A(t^n) \tag{9}$$

Now let us define,

$$\begin{aligned} C = C(i,j) = \frac{1}{N}U^T\left(x,t^i\right)U\left(x,t^j\right) \quad i,j = 1,2,3,\ldots,N \\ A = A^n = A\left(t^n\right) \quad n = 1,2,3,\ldots,N \end{aligned} \tag{10}$$

Hence, equation (9) can be written as:

$$CA = \lambda A \tag{11}$$

By solving the Eigen value problem presented by equation (11) $N$ mutually orthogonal Eigen Vectors $A_i, i = 1,2,3,\ldots,N$ are obtained, where the normalized POD modes are given as:

$$\emptyset_i(x) = \frac{\sum_{n=1}^{N} A_i(t^n)U(x,t^n)}{\left\|\sum_{n=1}^{N} A_i(t^n)U(x,t^n)\right\|} \tag{12}$$

In the present study, both the enstrophy and energy based POD analyses of a 2D plane along the centerline have been performed. For enstrophy based POD, the fluctuating components of vorticity vector are considered as the components of the auto correlation matrix, and the inner product of the field is given by the following equation:

$$U = \begin{bmatrix} \omega_x' & \omega_y' & \omega_z' \end{bmatrix}^T$$

$$\left(U(\boldsymbol{x},t^i), U(\boldsymbol{x},t^j)\right) = \int_\Omega \left\{ \omega_x'(\boldsymbol{x},t^i) \cdot \omega_x'(\boldsymbol{x},t^j) + \omega_y'(\boldsymbol{x},t^i) \cdot \omega_y'(\boldsymbol{x},t^j) + \omega_z'(\boldsymbol{x},t^i) \cdot \omega_z'(\boldsymbol{x},t^j) \right\} d\boldsymbol{x} \quad (13)$$

For energy based POD, the choice of variables become equally important for modal decomposition as the choices may vary depending on the nature of flow field, i.e. inclusion of variables for decomposition for incompressible and compressible cases may have significant effect on the output. Since the present case is compressible the choice of variables is no longer trivial. To represent the total energy of the flow accurately one must consider including thermodynamic variable to account for the internal energy and in the present study temperature is included such that the $U$ vector and its inner product can be given as:

$$U = \begin{bmatrix} u' & v' & w' & T' \end{bmatrix}^T$$

$$\left(U(\boldsymbol{x},t^i), U(\boldsymbol{x},t^j)\right) = \int_\Omega \left\{ u'(\boldsymbol{x},t^i) \cdot u'(\boldsymbol{x},t^j) + v'(\boldsymbol{x},t^i) \cdot v'(\boldsymbol{x},t^j) + w'(\boldsymbol{x},t^i) \cdot w'(\boldsymbol{x},t^j) + \gamma T'(\boldsymbol{x},t^i) \cdot T'(\boldsymbol{x},t^j) \right\} d\boldsymbol{x} \quad (14)$$

Here $\gamma$ is a scaling factor used to balance the velocity and temperature fluctuation energies. The optimal value of $\gamma$ proposed by Lumley & Poje[26] takes in to account the total energy from both the fluctuating field in a composite manner, Hasan & Sanghi[27] utilized this definition in their study related to thermally driven rotating cylinder. The scaling factor is defined as:

$$\gamma = \frac{\overline{\int_\Omega \left\{ u'(\boldsymbol{x},t^i) \cdot u'(\boldsymbol{x},t^i) + v'(\boldsymbol{x},t^i) \cdot v'(\boldsymbol{x},t^i) + w'(\boldsymbol{x},t^i) \cdot w'(\boldsymbol{x},t^i) \right\} d\boldsymbol{x}}}{\overline{\int_\Omega \left\{ T'(\boldsymbol{x},t^i) \cdot T'(\boldsymbol{x},t^i) \right\} d\boldsymbol{x}}} \quad (15)$$

where '‾' denotes averaging over time. Further details regarding the computation of POD modes and reconstruction can be found in the published literature [18-22].

**C. Dynamic Mode Decomposition**

Dynamic mode decomposition is a mathematical technique that gives overview of given temporal process with reduced dimension. Unlike POD dynamic mode decomposition sorts the dynamics flow by pure frequency thus revealing coherent structures orthogonal in temporal sense. For a given time sequence of data, $N$ number of snapshots equally separated in time ($\Delta t$) can be arranged in column vector as:

$$U_1^N = \{u_1, u_2, ..., u_N\} \tag{16}$$

Where, it is assumed that *A* is a matrix that linearly maps the $i_{th}$ sequence to the $(i + 1)_{th}$ field as,

$$u_{i+1} = Au_i \tag{15}$$

The Eigen values and vectors of matrix *A* describe the dynamic characteristics of the sequence $U_1^N$. As the number of snapshots is increased linear dependence is established and any further addition of snapshots does not improve the quality of observation. Hence, the vector $\underline{u_N}$ can be represented as liner combination of previous vectors as:

$$u_N = a_1 u_1 + a_2 u_2 + ... + a_{N-1} u_{N-1} + r \tag{16}$$

Here $a = a_1, a_2, \ldots, a_{N-1}$ represents the coefficients of DMD, *r* is the residual vector, in matrix form it can be written as

$$u_n = U_1^{N-1} a + r \tag{17}$$

Following the approach of Schmid[23], the above equation can be recast as,

$$AU_1^{N-1} = U_2^N = U_1^{N-1} S + r.e_{N-1}^T \tag{18}$$

Here *S* is an unknown matrix of companion type and $e_{N-1}$ is the $(N-1)_{th}$ unit vector. To determine the *S* matrix, minimization of *r* is carried out in the least square sense which is further solved though the *QR* factorization. The Eigen values and Eigen vectors of *S* then represent the dynamic modes of the flow field under consideration. More details can be found in the literature by Schmid[23].

**D. Computational Details**

    The computational domain involves a channel within which strut is placed as can be seen from Fig 1, the coordinate originates at the tip of the wedge. At the end of the strut, i.e. in the base region hydrogen is injected into the oncoming supersonic flow. Fig. 1 exhibits the details of two struts namely, Straight & Tapered configurations, which are investigated to accommodate the effect of recirculation strength on the mixing and jet spreading. Since the channel length is several order larger compared to the jet diameter (0.6 & 1 mm), multiblock meshing philosophy is adopted to maintain the optimum grid

size. The three set of grids are generated for *TS−0.6* case namely G1, G2 & G3 to study the sensibility of numeric toward the spatial discretization further detail is presented in Table II.

Table II. Meshing parameter for all the grids. $\Delta y_w$ and $\Delta y_{jet}$ denotes the grid spacing along the transverse direction in near wall and jet region. $N_x$, $N_y$ and $N_z$ denote the grid point distribution along the streamwise, transverse and spanwise direction respectively and $D_j$ (0.6 mm) signifies jet height

|                 | G1        | G2         | G3          |
|-----------------|-----------|------------|-------------|
| $\Delta y_w$    | $0.8D_j$  | $0.2D_j$   | $0.1D_j$    |
| $\Delta y_{jet}$| $0.35D_j$ | $0.08D_j$  | $0.072D_j$  |
| $N_x$           | 730       | 1105       | 1326        |
| $N_y$           | 142       | 210        | 250         |
| $N_z$           | 18        | 25         | 30          |

The jet region including the strut boundary is resolved appropriately to capture the evolution of shear layer and recirculation region effectively. Also in the near field region, enough resolution along the streamwise and spanwise directions is ensured to capture the wave system efficiently. For all the cases top and bottom walls are modeled ($y+ \approx 30$) because those walls may have very little or no influence on the mixing characteristics, especially in the near-field region. The wall modeling approach adopted here is already validated and details can be found in Ref.[18]. The details regarding the test cases are presented in Table III. Table IV presents the inlet boundary condition utilized in the present computation. Dirichlet boundary condition is invoked for both air and hydrogen inlet, at the outlet non-reflecting boundary condition is prescribed to avoid wave reflection. One may refer to the work of Soni at el.[18] for further details regarding the implementation of the non-reflecting boundary condition and wall modeling approach. On the top, bottom and strut region adiabatic no-slip boundary condition is imposed and periodic boundary condition along the spanwise direction is imposed. The significant difference in the inlet velocity for air and hydrogen is due to the gas constant & specific heat ratios.

TABLE III. Details of the test cases investigated. (subscript $\infty$ & $j$ refer to freestream and jet freestream conditions)

|    |     | $M_j$ | $D_j$ (mm) | $M_\infty$ |
|----|-----|-------|------------|------------|
| SS |     | 2.3   | 0.6 & 1    | 2          |
|    |     | 1     | 0.6        | 2          |
| TS |     | 2.3   | 0.6 & 1    | 2          |
|    |     | 1     | 0.6        | 2          |

TABLE IV. Inlet condition for all the cases. ($U_\infty = 2203$ & $958$ m/s correspond to $M_j = 2.3$ & $1$ respectively. M, P, U and T represent the Mach number, pressure, streamwise velocity and temperature respectively.)

| Parameter | Air | Hydrogen |
| --- | --- | --- |
| $P_\infty (Pa)$ | 49.5 | 29.5 |
| $T_\infty (K)$ | 159 | 151 |
| $M_\infty$ | 2 | 2.3 & 1 |
| $U_\infty$ (m/s) | 505 | 2203 & 958 |
| $Y_{N_2}$ | 0.76699 | 0 |
| $Y_{O_2}$ | 0.23301 | 0 |
| $Y_{H_2}$ | 0 | 1 |

The calculation is performed in the OpenFOAM framework, the density based solver which is already validated[18, 28] and modified to solve for species transport equation along with mass, momentum, and energy. The solver utilizes central scheme which is an alternative approach to Riemann solver offering an accurate non-oscillatory solution. More detailed information about the implementation in OpenFOAM can be found in Greenshields et al. [29] and Kurganov & Tadmor[30]. The dynamic sgs model invoked in the current computation is based on the two model constants which are evaluated dynamically[31]. In present simulation, second order backward Euler scheme is utilized for the time integration whereas viscid and inviscid fluxes are discretized using central difference and TVD scheme respectively. The parallel processing is achieved through the message passing interface (MPI) technique. The statistics are collected for the simulations over ~30 non-dimensional times while maintaining CFL number below 0.5.

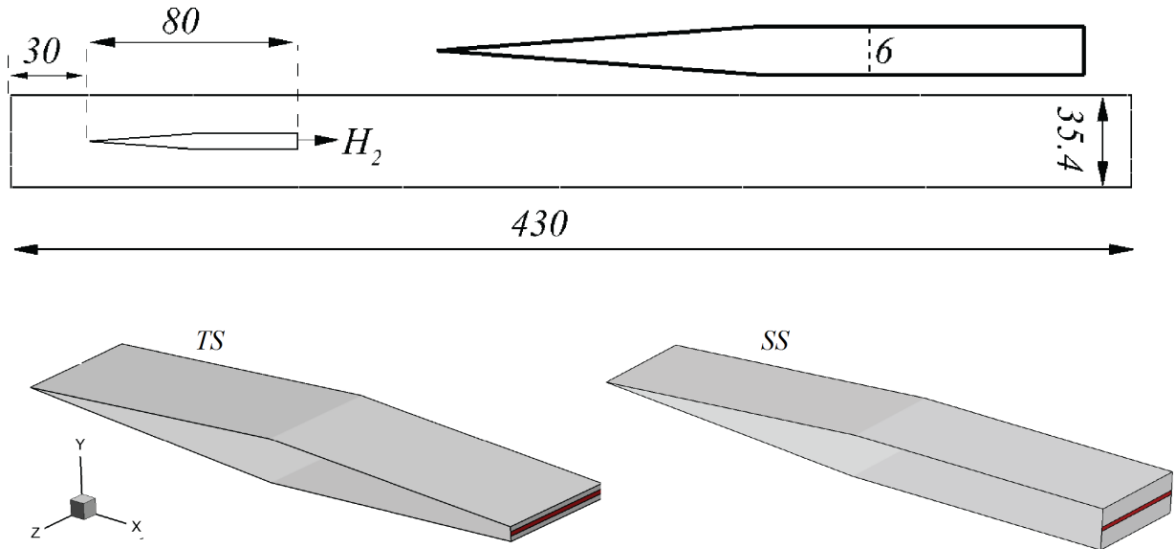

FIG. 1. Details of computational domain both the strut geometry are shown; all the dimensions are in mm

## III. RESULTS & DISCUSSION

### A. Mean and Instantaneous data analysis

Initially, the grid independence study is carried out using three grids G1, G2, and G3. The hydrogen and nitrogen mole fractions (Fig. 2) trend at different streamwise locations for all the grids are compared with the experimental results. It reveals that G2 & G3 follow the experimental observation closely and G1 suffers significantly as one move towards the downstream region. Also time averaged streamwise velocity and Reynolds stress component is presented in Fig. 3 for all three grids at different axial locations downstream of the injector. The result of G1 differs significantly from G2 and G3, and the under-prediction in transverse direction clearly points out the inability of G1 grid to efficiently capture the physical phenomena. Hence, upon combining the information, it is worth noticing that grids G2 & G3 perform better and the results are in excellent agreement with each other. Henceforth all the results presented correspond to the grid G2 for detailed analysis. To gain more confidence on the chosen grid, index of the grid quality, proposed by the Celik[32], which is based on the eddy viscosity ratio, is recast as,

$$LES_{IQ} = \frac{1}{1 + \alpha_v \left(\frac{\mu_t}{\mu_{eff}}\right)^n} \tag{19}$$

where, n = 0.53, $\alpha_v$ = 0.05, $\mu_t$ and $\mu_{eff}$ are eddy and effective viscosity respectively. They suggested that $LES_{IQ}$ of 75% to 80% is acceptable for most of the problems of engineering interest usually occurring at high Reynolds number. The $LES_{IQ}$ presented in Fig. 4 also suggests sufficient grid resolution over the whole domain with $LES_{IQ}$ well above 90%.

To quantify the mixing effectiveness for different configuration mixing layer thickness, which is an important parameter to get insight into the extent of diffusion in the transverse direction, is plotted along the streamwise direction and depicted in Fig. 5. Here, the mixing layer thickness is defined based on the distance from jet centerline to the point in the transverse direction where mole fraction of hydrogen reduces to 5% of the injected value. Similar kind of definition to assess the mixing layer growth was also utilized in the study of Gerlinger & Bruggemann[5] and Ben-Yakar et al.[33]. It is interesting to note that the straight strut penetrates more in the lateral direction which is true for both the convective Mach numbers but more spreading occurs for the $M_c$-1.4 case. This could be attributed to the shear strength (R) across the layer; means the higher shear induces relatively larger structures along the shear layer. It is already reported in the literature that the larger velocity difference between the freestream fluid and injectant leads to the enhanced mixing across the shear layer through the stretching/tearing mechanism[34].

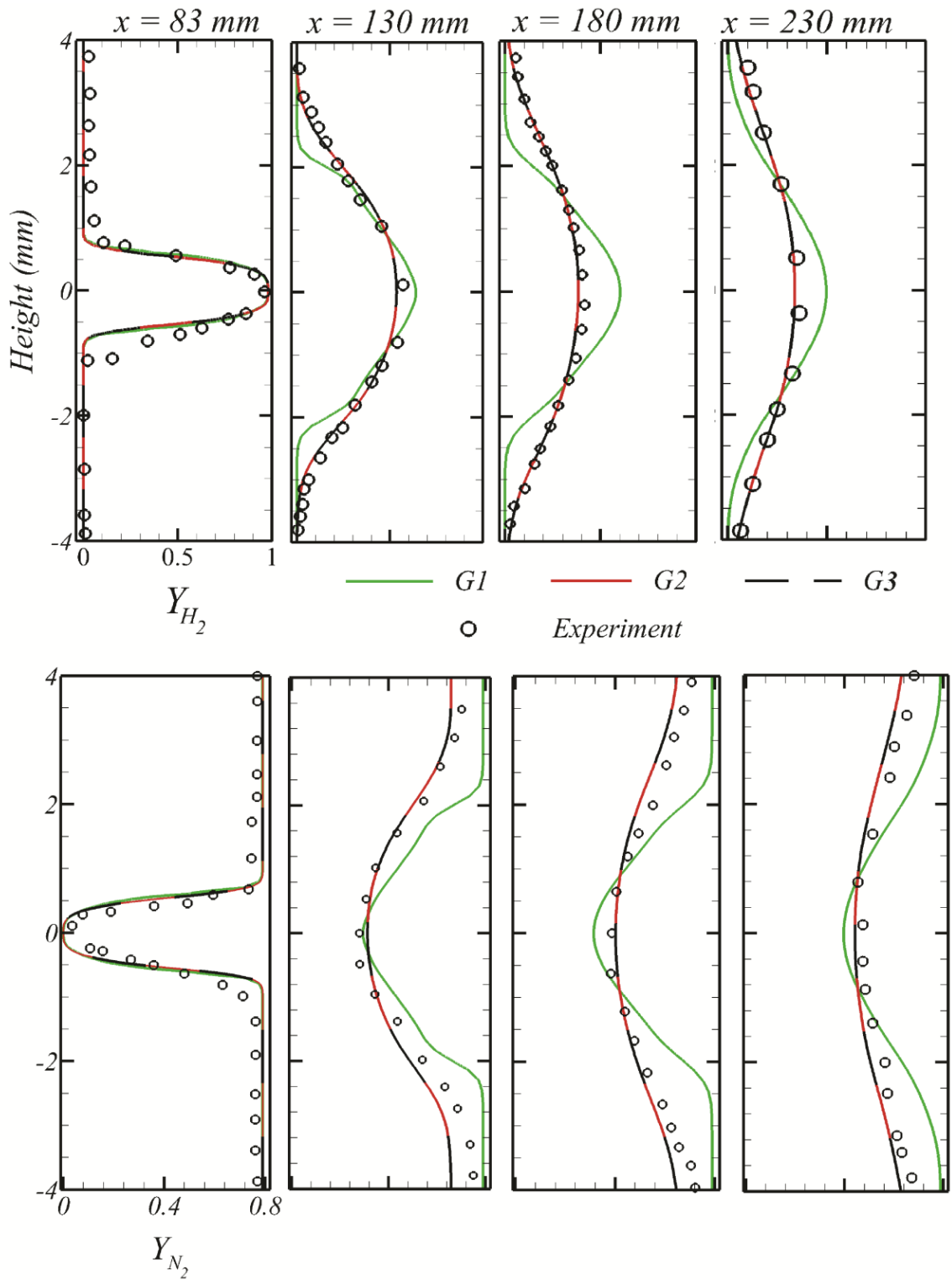

FIG. 2. Time-averaged mole fraction of hydrogen (top) & nitrogen (bottom) at different axial location for TS − 0.6 Case at ($M_c = 1.4$)

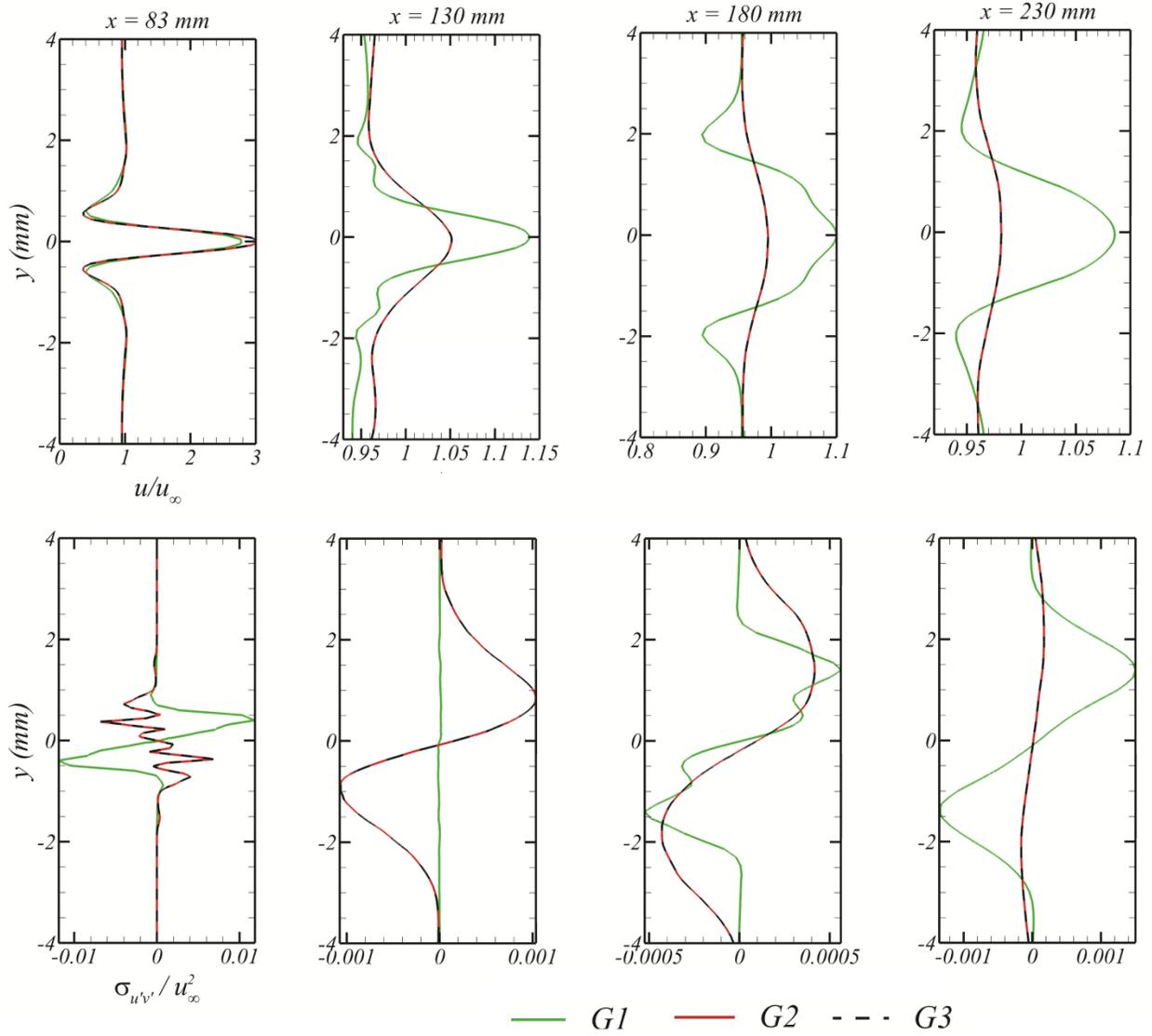

FIG. 3. Normalized mean streamwise velocity and Reynolds stress comparison along different axial location for TS − 0.6 Case for various grids at $M_c = 1.4$

In the present investigation, it becomes obvious that $M_c$-1.4 cases are subjected to larger velocity gradients and the effect of which is manifested in the shear layer growth and mixing efficiency (Fig. 5). This explains the reason for the enhanced performance of the SS configuration over TS cases however detailed discussion and inspection will be the subject of subsequent sections. When comparing TS cases for both the jet thickness (0.6 & 1 mm), it appears 1mm cases perform better which is primarily due to the variation in the strength of the recirculation zone, thereby revealing the fact that there could be multiple phenomena responsible for this type of behavior as discussed later.

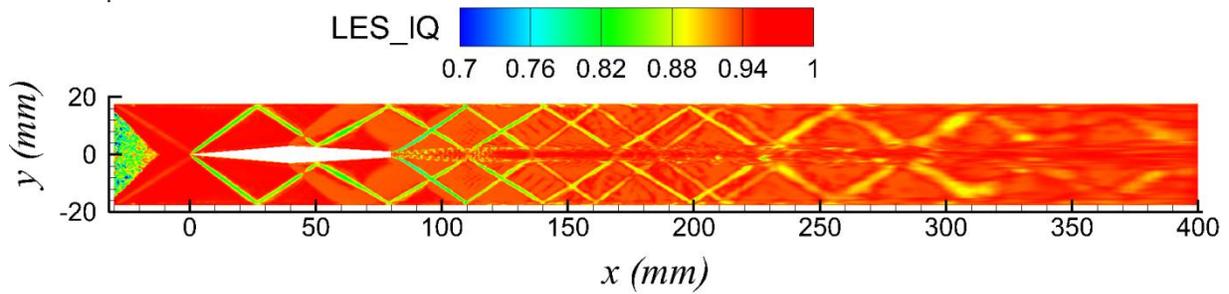

FIG. 4. LES_IQ to demonstrate the grid resolution of Grid G2

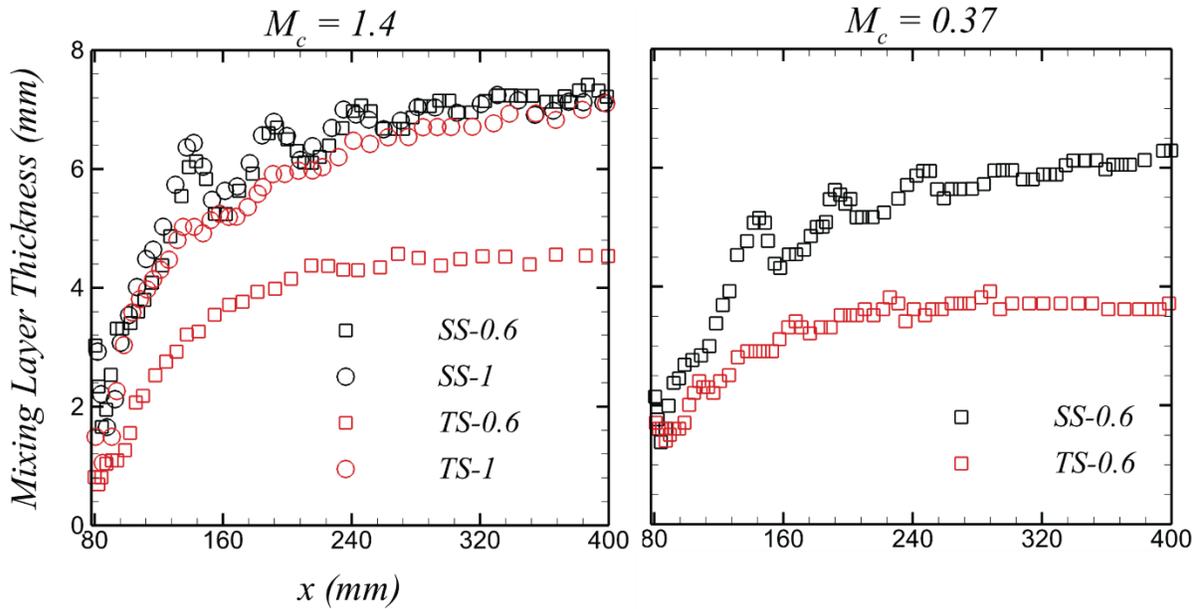

FIG. 5. Spatial growth of mixing layer thickness for both convective Mach number

Upon inspecting the trend of mixing layer thickness, one can notice the initial decrease in thickness due to the presence of the expansion fan at the edge of the base region, beyond which the rapid increase is witnessed for all the cases. This happens as the turbulence diffusion escalates and dominates the mixing process in this region ($X > 240$ mm). However, at the edge of the high diffusive region, the rise in thickness becomes less steep. The wavy pattern is more dominant for the straight strut configuration as this geometry experiences strong shock train whereas the splitter plate configuration exhibited much smoother profile in the absence of strong shock train.

To further quantify the mixing and complement the observation of earlier sections, mixing efficiency ($\eta_m$) is calculated. Mixing efficiency is that fraction of least available reactant that can react if the mixture is brought to chemical equilibrium. The two-part definition of mixing efficiency is defined as [Mao et al.[35]]

$$\eta_m = \frac{\int Y_f \rho u \, dA}{\int Y \rho u \, dA} \qquad (20)$$

where,

$$Y_f = \begin{cases} Y, & Y \leq Y_s \\ Y_s(1-Y)/(1-Y_s), & Y > Y_s \end{cases}$$

where $Y_f$ is the fraction of hydrogen mixed in the proportion that can readily react, $Y$ is the hydrogen mass fraction and $Y_s = 0.0292$ is the stoichiometric mass fraction of hydrogen. The integral is performed over cross-sectional area perpendicular to the flow direction. The mixing efficiency is reported in Fig. 6 which clearly suggests the superiority of straight configuration that offers higher mixing compared to the tapered strut. The $\eta_m$ for *SS−0.6* case shows steep rise with *100%* mixing at about *X = 280 mm*, whereas *TS−0.6* case achieves similar efficiency at the channel end. A similar trend can be witnessed for the *1 mm* jet height case, once again straight strut achieves *100%* efficiency but at slightly downstream location compared to the *SS-0.6* case. The strength of the recirculation region remains the primary contributor to the significant difference in the η$_m$ curve or both *TS-0.6* and *TS-1* cases.

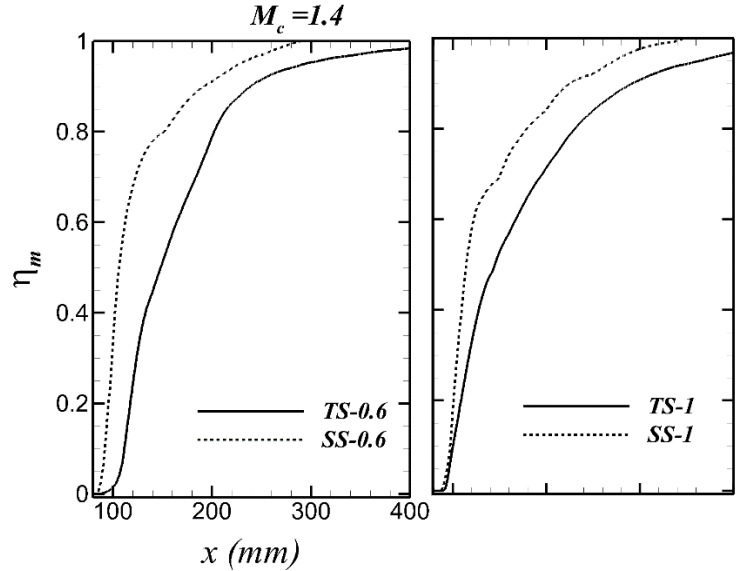

FIG. 6. Spatial variation of mixing efficiency along the jet centerline

The above discussion leads to an important question regarding the parameters that affect the flow field and dictate the mixing process. Is convective Mach number a sufficient parameter to evaluate shear layer growth effectively for complex configurations? Or are there other parameters that need to be analyzed? Hence, it is quite essential to consider the parameters such as the turbulence generated close to the injector, strength of recirculation region and larger velocity gradient along the lateral direction, for better understanding and quantification of mixing characteristics. In literature, some of these parameters were considered to quantify mixing in the context of simple splitter plate arrangement. However, due to geometric complexity in the present case, the factors that affect the mixing are completely different and we have attempted to analyze the same through our simulations and modal analysis as well.

To understand the effect of shear strength and its consequence on turbulence production, the normal gradient of streamwise velocity is presented at two downstream locations in Fig. 7. A significant difference in the peak value is observed close to the injector, while at $X = 83$ mm higher gradient is found to be present and the gradient vanishes to approximately 1% within 50 mm from the injector. This observation strongly suggests that the effect of convective Mach number is very much limited only in near injector region. Due to this, the flow close to injector region is dominated by the compressibility effects and flow behavior similar to pure shear flow is witnessed. One of the most important characteristics of pure shear flow is being the reduction in mixing layer thickness with increasing convective Mach number. The same can be verified from the Fig. 5 wherein the vicinity of injector mixing layer thickness reduces before rising again, this reduction is more significant for $M_c = 1.4$ cases compared to $M_c = 0.37$. The high gradient along shear layer leads to the rolling up of shear layer and the two-dimensional rollers are formed which undergo pairing/merging and stretching as they convect downstream. Flow subjected to higher gradient and turbulence intensity experiences more turbulence production which is beneficial for this type of flow. Higher turbulence production across shear layer is known to dictate the turbulent mixing effect of which is clearly visible from Fig. 6. So this suggests that convective Mach number could be an important parameter but not sufficient, especially where the pure shear flow is not present as depicted in the present study.

Moving forward with our discussion, the streamwise and lateral turbulence intensities are presented in Fig. 8 at two axial locations. The figure reveals a significant difference in fluctuation level with higher being at $M_c = 1.4$, further downstream the intensity is reduced with more lateral extent which can be attributed to the exchange of the fluid between these two regions. This explains that with increasing downstream region more of the thickness of the shear layer is turbulent which is true for both the cases. The increased shear induces the increased instability and promotes the higher jet spreading rate. These findings are in contrast to the results of Elliot & Samimy[36], as they reported a reduction in turbulent fluctuation with increasing convective Mach number for splitter plate configuration.

The Reynolds stress is presented for both $M_c$, the extremes in the plot indicate the peak turbulence level which occurs at the shear layer along the centerline. A similar trend is observed for both Mach numbers with a large difference in the highest values. On combining this information one can certainly notice that the maximum turbulence level occurs for $M_c = 1.4$ shear layer and the effect of turbulence is manifested in the mixing layer thickness through the presence of large (coherent) scale

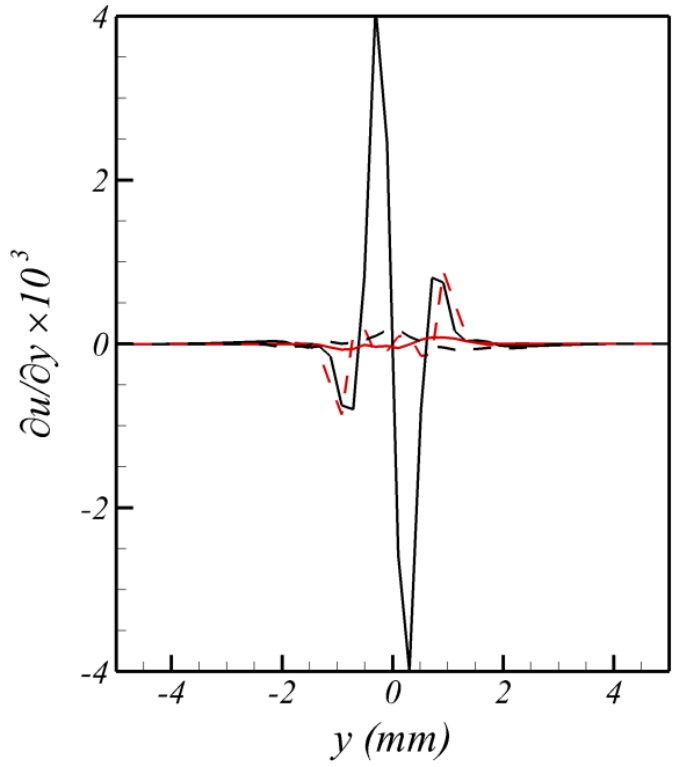

FIG 7. Velocity gradient plot for both the convective Mach numbers (Red line represents $M_c = 0.37$ and Black line corresponds to $M_c = 1.4$, solid and dashed lines represents axial location $x = 83$ & $130$ mm respectively)

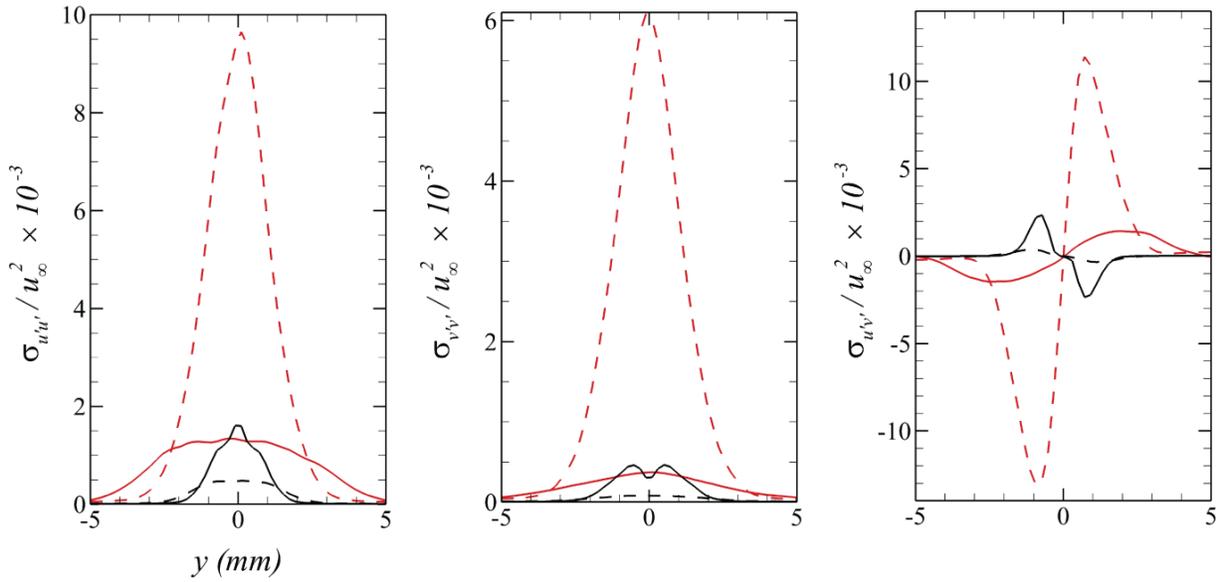

FIG. 8. Streamwise & lateral turbulence intensity and Reynold stress component (*Red and Black line correspond to $M_c = 1.4$ & $0.37$ respectively, solid and dashed line correspond to $x = 130$ & $230$ mm*)

structures. The large-scale structures transport the fluid between the outer and inner region hence it becomes vital to understand the behavior and dynamics of these coherent vortices. Therefore, now onwards, our particular focus is on the characterization of the large-scale structures and its effect on the overall flow field with special attention on mixing behavior.

**B. Characterization of large-scale structures**

In Fig. 9 (A) & (B) the gradient of density for both geometric configuration are presented for $M_c$ =1.4 & 0.37 to appreciate the complex nature of the flow downstream of the injectors. In Fig. 8(A) some of the important flow features are labeled, the complex interaction of wave system is observed with strong reattachment shock, where the circle highlights the behavior of structures over spatial extent. As the vortices are convected downstream the effect of diffusion is manifested in the enhanced scale of the vortices, and the same is true for the *TS* cases. However, the huge difference between the spatial scales of structures is observed between both the configurations. It can be seen that the shock train appears to be steady and is not affected by the presence of the recirculation zone. However, the recirculating region created at the strut base on either side confined between the high-speed jet and free shear layer introduces unsteadiness. The recirculation region is subjected to high shear and this leads to the formation of vortical structures. These unsteady vortical structures aid in mixing of hydrogen jet. Further downstream, the interaction of jet and shear layer leads to vortex breakdown with the presence of local vortical structures which can be clearly noticed from the Fig. 9, where various vortical structures along the shear layer are visible. Close to the injector the jet breakdown is triggered due to the $K-H$ instabilities and leads to the formation of spanwise rollers which are of two-dimensional nature in near-field but in far-field higher dimensional structures are also present.

Further the effect of recirculation zone is significant on the structure and growth of the shear layer, as seen in Fig. 9. For cases involving *SS* configuration more chaotic structures are present which appear to be spread further outward as opposed to the *TS* configuration. It is clear that the vortical structures in the vicinity of the injector are largely two-dimensional in nature and mimics spanwise rollers as observed by Soni et al.[18] in their investigation on flow past supersonic backward-facing step. However, further downstream more three-dimensional structures are present due to high shear across the shear layer, increasing the mass transfer and thus aiding in mixing. Additionally, the shock/shear and crossing shock lead to further enhancement in vortex breakdown through the baroclinic torque mechanism. The early breakdown of the vortices in the wake region of the strut offers better mixing by increased diffusion leading to the entrainment of external fluid along the jet.

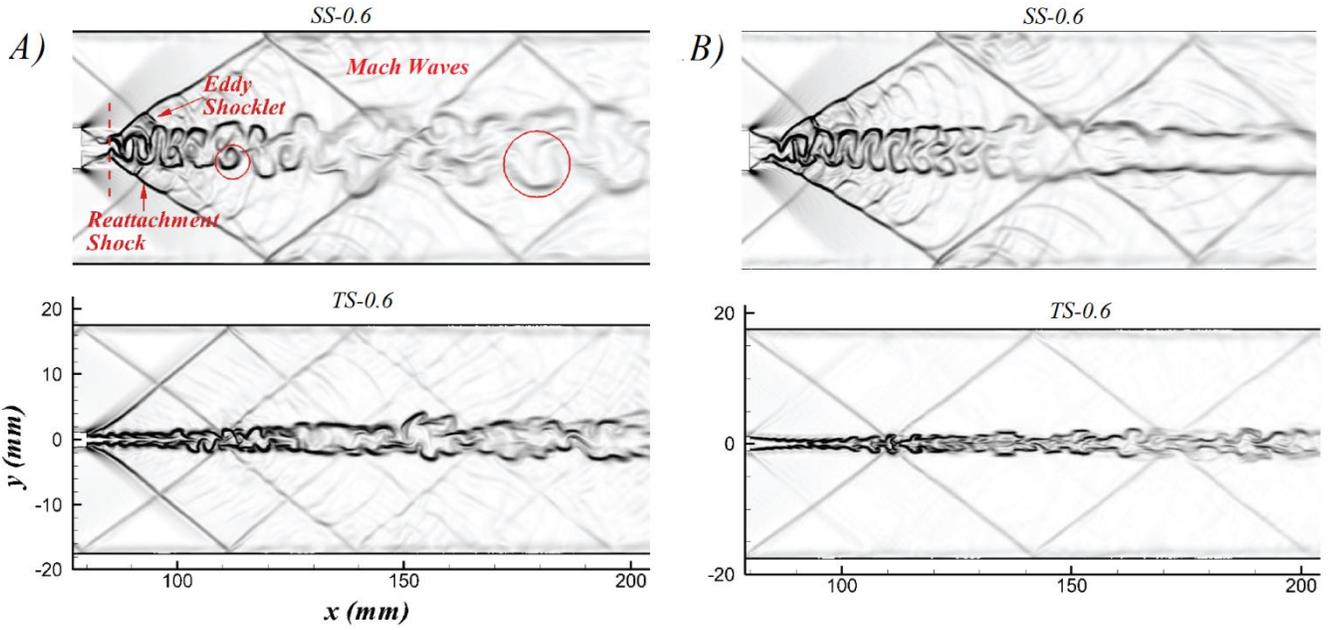

FIG. 9. Instantaneous density gradient contour for A) $M_c=1.4$ & B) $M_c= 0.37$

Another factor that is more intriguing is the striking difference in the *SS* & *TS* cases for both the convective Mach number (Fig. 9). A closer look reveals that there is early onset of vortex breakdown for SS configuration; qualitative inspection reveals that for *SS* cases it is around $X \approx 83-85$ mm, whereas for *TS* configuration it is around $X \approx 105-110$ mm. It can be expected that early onset of vortex breakdown and presence of stronger shock would result in improved mixing due to increased mass transfer across the shear layer. Also, it appears that there is a periodic vortex shedding past the breakup point which suggests shear layer flapping phenomenon. Another factor that is worth analyzing is the second invariant of velocity tensor, which is presented in Fig. 10 and colored with the instantaneous hydrogen mass fraction, exhibits the presence of vortical structures present downstream of the injection point. Various vortical structures are found to be present in the walls, shear layers and in the wake region. In the immediate vicinity of the jet, most of the vortical structures are generated due to the *K−H* instability owing to the interaction of recirculation region and the shear layer. After the occurrence of reattachment shock, vortex breakup initiates and leads to the formation of 3-dimensional structures at the downstream. Other than these vortices, structures generated due to oblique shock wave and expansion fan are also visible.

At the downstream of the strut, wake region appears to start spreading but due to the presence of reattachment shock again converges toward the jet centerline forming neck before spreading out again. The similar phenomenon is observed in the Fig. 9, while the current observation is consistent with the results of Burns et al.[34]. Along the shear layer the fine scale structures are present; however, past recompression point, larger structures are visible in the wake region. Comparing *SS−0.6* with the *TS−0.6* case, much larger rollers are visible and the pairing/merging of vortices past breakdown also becomes apparent for the *SS-0.6* case. Apart from near jet region vortices, recirculating structures are also present on the walls which are mostly due to the shock/boundary layer interaction however as discussed earlier these vortices play little or no role in the mixing process.

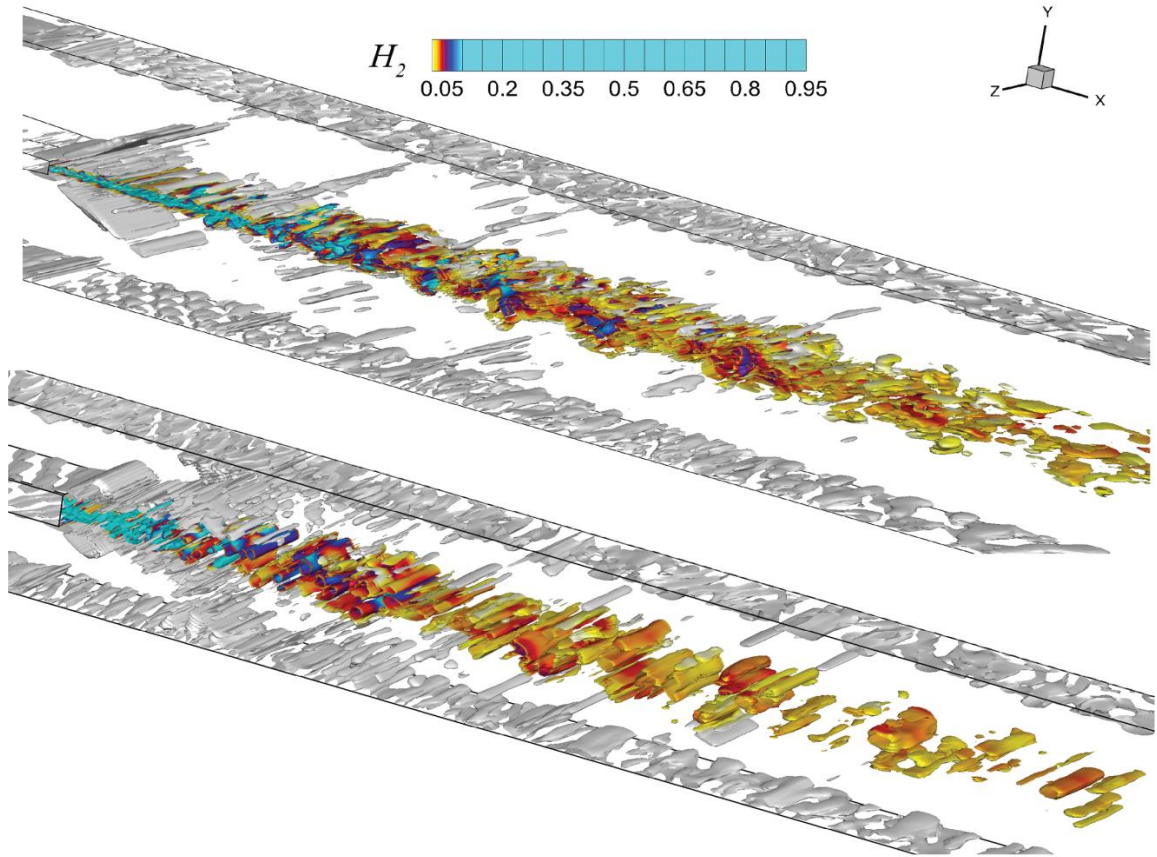

FIG. 10. Second Invariant of velocity tensor for *TS-0.6* case (Top) and *SS-0.6* (Bottom) for $M_c = 1.4$

Instantaneous vorticity contour for *SS & TS* configuration with a jet height of 0.6 mm is plotted in Fig. 11, where the isoline corresponds to the constant instantaneous hydrogen mass fraction of 0.002. Comparing both the cases it is found that the *SS* configuration is dominated by higher spreading rate. Although there is a similarity in the profile, spreading rate is significantly higher in Fig. 9(A). Gutmark et al.[8] investigated a range of convective Mach numbers and concluded that the spreading rate increases with the increase in shear, where the shear strength is characterized by the ratio of two velocities ($R = u_2/u_1$). This is due to the difference in the lip thickness which, on the other hand, induces the recirculation zone of different strengths as previously reported by Gerlinger & Bruggemann[5] and Soni & De[25]. It is noteworthy to mention that the *SS* configuration induces stronger (& larger) recirculation bubble in the strut bases region and hence the flow experiences larger gradient across the low & high-speed side.

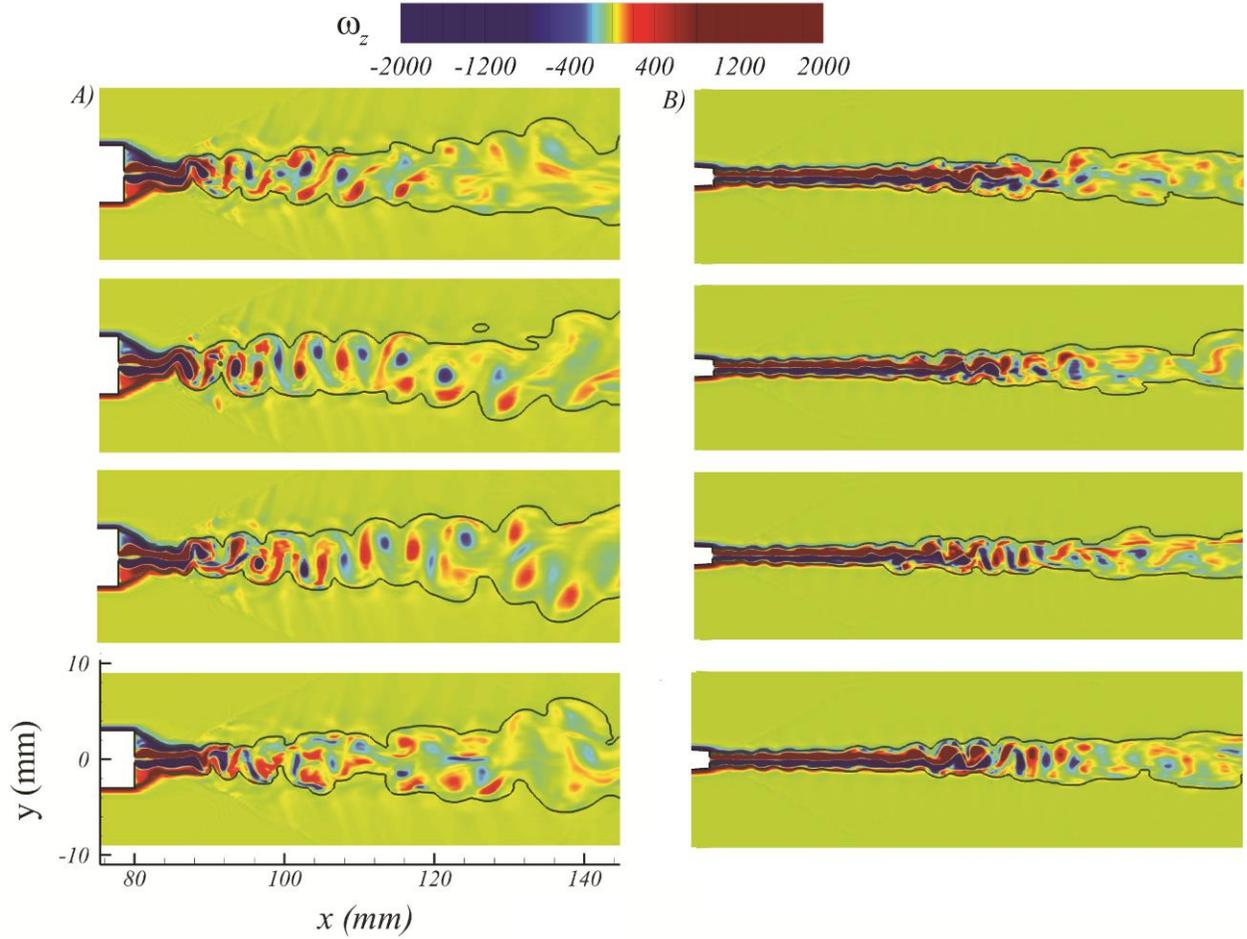

FIG. 11 Instantaneous vorticity contour of A) *SS-0.6* & B) *TS − 0.6 for $M_c$-1.4*, isoline depicts the 0.2% of hydrogen mole fraction

The vortex breakdown for *SS* case occurs through the recompression shock at $X \approx 84$ mm whereas in case of the *TS-0.6* case the jet breakup does not initiate in a similar fashion instead it undergoes compression through the crossing wave system. This could be due to the presence of stronger expansion fan at the middle of strut geometry which leads to acceleration of flow due to the tapered configuration. Also at the strut base, lip thickness being small for *TS* cases recompression shock strength suffers significantly when compared to *SS* configurations. For *TS−0.6* the wake starts to grow at $X \approx 110$ mm before that the *K−H* instability is clearly visible along the shear layer. Fig. 10 also points toward the vortical shedding past the vortex breakdown point, the behavior is quite similar to vortex Street shedding past cylinder. The formation of large-scale vortical structures of the alternating sign can be seen past this point; it appears to be well organized and periodic in nature. The wake region can also be found to be broadening as one move downstream, and the size and spacing of vortical structures increase with the increase in the wake width. The axial vorticity generated close to injector leads to a lifting of fuel jet from the surface and enhances mixing, as depicted in Fig. 6.

There are different ways to identify the physics behind the vortex breakdown as suggested by previous researchers in literature. Hiejima[37, 38 and 39] reported that shock wave interacting with vortices may trigger the breakdown. Following to their work, Delery[40] and Kalkhoran & Smart[41] also observed that the onset of breakdown depends on the intensity & angle of impinging the shock wave, circulation, and axial velocity deficit. They also pointed out the reason behind the vortex break down is mainly due to the interaction between a vortex and oblique shock wave, leading to the localized subsonic region; as suggested by the Hiejima[37] in their study pertaining to supersonic combustion in hyper mixer strut configuration. It was also discerned that these localized pockets of subsonic flow are ideal for flame holding. However in the present investigation, the flow never decelerates to lower subsonic value, rather the Mach number in the range of 0.9–1 is observed at downstream of the strut. The presence of lower subsonic Mach number in reacting flows is associated with the temperature rise due to combustion, which is not present in the current calculations. Moreover, the location of vortex breakdown can also be quantified by integrating enstrophy over a plane perpendicular to streamwise direction as proposed by Pierro & Abid[42]. The presence of local maxima in an axial variation of integrated enstrophy indicates the vortex breakdown location. The enstrophy is calculated as:

$$\varepsilon = \int_\Omega \frac{1}{2} \omega_i \omega_i \, d\Omega \qquad (21)$$

$$\varepsilon_{y+z} = \varepsilon - \int_\Omega \frac{1}{2} \omega_x^2 \, d\Omega \qquad (22)$$

Fig. 12 presents the $\varepsilon_{y+z}$ calculated by subtracting streamwise vorticity contribution for all cases past the injection point. The variation is presented only in the near field region as enstrophy diminishes rapidly. The presence of local maxima in the plot suggests the vortex breakdown locations, which is consistent with the observations of Fig. 9 & 11. At these locations, vortices undergo tremendous stretching which upon convecting downstream entrain fluid from the outer layer and vice versa. For both the convective Mach numbers, *SS* configuration has early onset of the vortex breakdown due to which the mixing layer grows rapidly in an outward direction, thereby exhibiting better mixing characteristics (Fig. 6). For these instantaneous analyses, it is understood that the vertical structures along with their interaction with the shock structures are primarily responsible for the mixing characteristics as observed in Fig. 6. However, this analysis is not sufficient to identify the vortex structures and the energy mode associated with the shock-vortex interactions. Hence, we extend our analysis (following subsections) using the modal decomposition of the flow field to quantify the same.

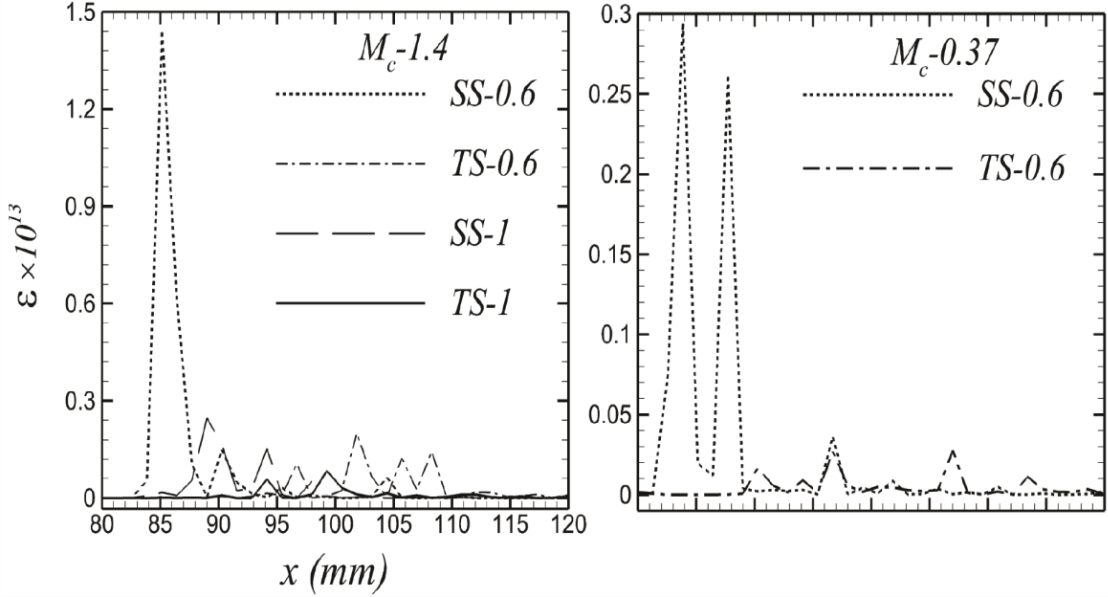

FIG. 12. Spatial variation of enstrophy for $M_c = 1.4$ & $0.37$ in the near field region

**C. Proper Orthogonal Decomposition**

The POD is a powerful tool that helps in identifying the coherence structure in the turbulent flows. The Eigen modes that are computed and arranged on the basis of energy content are representative of various coherent spatial features dominating the dynamics of flow. Generally, first few modes are enough to represent the flow physics, also the reconstruction with few modes are often sufficient to represent the original flow field. To demonstrate the convergence of energy composition, decomposition has been performed with three sets of snapshots equally spaced in time. Noteworthy to mention that the decomposition has been reported on the original computational grid (G2), means we have not interpolated the computational grid to any other grid for modal decomposition. Table V presents the relative Eigen values of first 5 Eigen modes for $N=80$, $100$ & $120$, which confirms that the variation in the energy content is minimal and hence the detailed discussion or analysis is reported using $N = 120$ only. The decomposition is performed on the $z=0$ plane (mid plane) and the relative Eigen value for $TS-0.6$ & $SS-0.6$ cases based on velocity (energy) and vorticity (enstrophy) based POD are presented in the Fig. 13 and the energy content for first 5 modes for both the approaches are reported in Table VI. For $SS-0.6$ case using energy based POD the first mode alone contributes toward the 97% energy, whereas for the $TS-0.6$ case it is 99% energy; and the contribution from remaining modes is found to be negligible. In case of vorticity based POD, the energy distribution for the first mode is a unity which corresponds to the instantaneous flow field.

Table V. First five Eigen values for the demonstration of snapshots convergence

| n = 80 | n = 100 | n = 120 |
|---|---|---|
| 0.9743 | 0.9743 | 0.9745 |
| 0.0041 | 0.0043 | 0.0044 |
| 0.0031 | 0.0031 | 0.0031 |
| 0.0021 | 0.0023 | 0.0023 |
| 0.0015 | 0.0016 | 0.0016 |

Table VI. Eigen values corresponding to first five dominant modes for both energy and vorticity based decomposition

| Modes | Eigenvalues (Energy POD) | | Eigenvalues (Vorticity POD) | |
|---|---|---|---|---|
| | TS-0.6 | SS-0.6 | TS-0.6 | SS-0.6 |
| 1 | 0.993 | 0.974 | 1 | 1 |
| 2 | 0.00041 | 0.0043 | 0.0062 | 0.0275 |
| 3 | 0.00037 | 0.0030 | 0.0032 | 0.0247 |
| 4 | 0.0003 | 0.0021 | 0.0024 | 0.0075 |
| 5 | 0.00028 | 0.0016 | 0.0023 | 0.0074 |

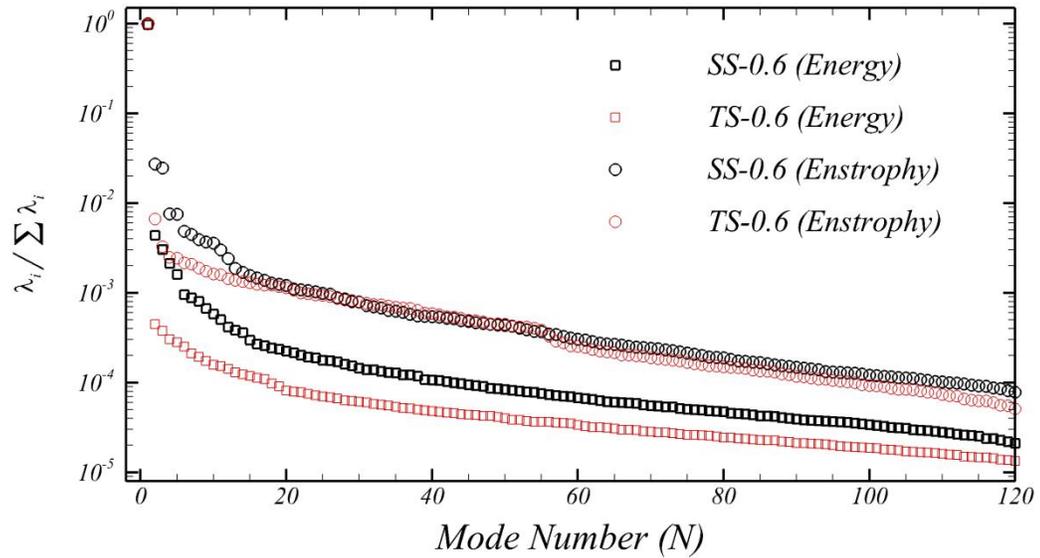

FIG. 13. Relative Eigen value distribution across the modes for velocity (energy) & vorticity based decomposition

Figure 14 presents the first five Eigen modes for $SS-0.6$ case ($M_c-1.4$), where the first mode can be thought of as representative of mean motion and the contour represents the transverse velocity. The recompression shock and trace of crossing shock wave is clearly captured in all the five modes. The extent of spreading can be witnessed in the presence of various vortical structures, which inherently affects the mixing characteristics. While comparing Fig. 14 and Figs. 9-11, a visible consistency in the pattern is observed where the small-scale structures are initially seen close to the injector located around the inner shear layer. These vortices represent the *K-H* structures which upon convecting downstream interact with the shock and breakdown. Past this point, the structures tend to grow where mostly small-scale structures are present along the inner shear layer and larger structures convecting downstream are slightly lifted and are present on the outer edge of the shear layer. These large-scale structures facilitate the exchange of the mass & momentum between the two regions. The mass from the inner region is ejected to the outer moving layer and from outer layer it is brought into the inner layer, thereby leading to the mixing of two streams. Eigen modes 2 to 5 represents fewer energy structures with some fine scale structures which are associated with the high-frequency turbulent vortices. Modes 2, 3 & 4 exhibit quite similar in behavior especially in the near field region where larger structures elongated in the transverse direction and their shapes are found to be elliptic in nature. Moving further downstream, the mixed scales are found to be present. Mode 5 clearly suggests the symmetric vortical distribution throughout the domain and hence can be associated to the periodic flapping of the shear layer. In fact, all the modes mostly exhibit symmetric behavior along the jet centerline, with some antisymmetric distribution in a few downstream regions.

One important take away from the Eigen modes contour plot is the extent of spreading; it seems that aggressive growth in mixing layer happens in the near field region beyond which the spreading becomes less steep. Here the domain is presented only till 200 mm in the streamwise direction as the major interest lies in analyzing the near-field mixing behavior. Most of the large-scale structures are present within this range (200mm) beyond which small-scale structures are witnessed which is due to the vortex stretching leading to breakdown. The largest structure noticed through the Eigen modes is approximately of the order of $8D_j-16D_j$ ($D_j=0.6$ mm) and the average large-scale structure is between $4D_j-6D_j$. This observation sheds light into the near field mixing behavior for the $SS-0.6$ case as witnessed in Fig. 6, where about 90% mixing efficiency is achieved at 200 mm.

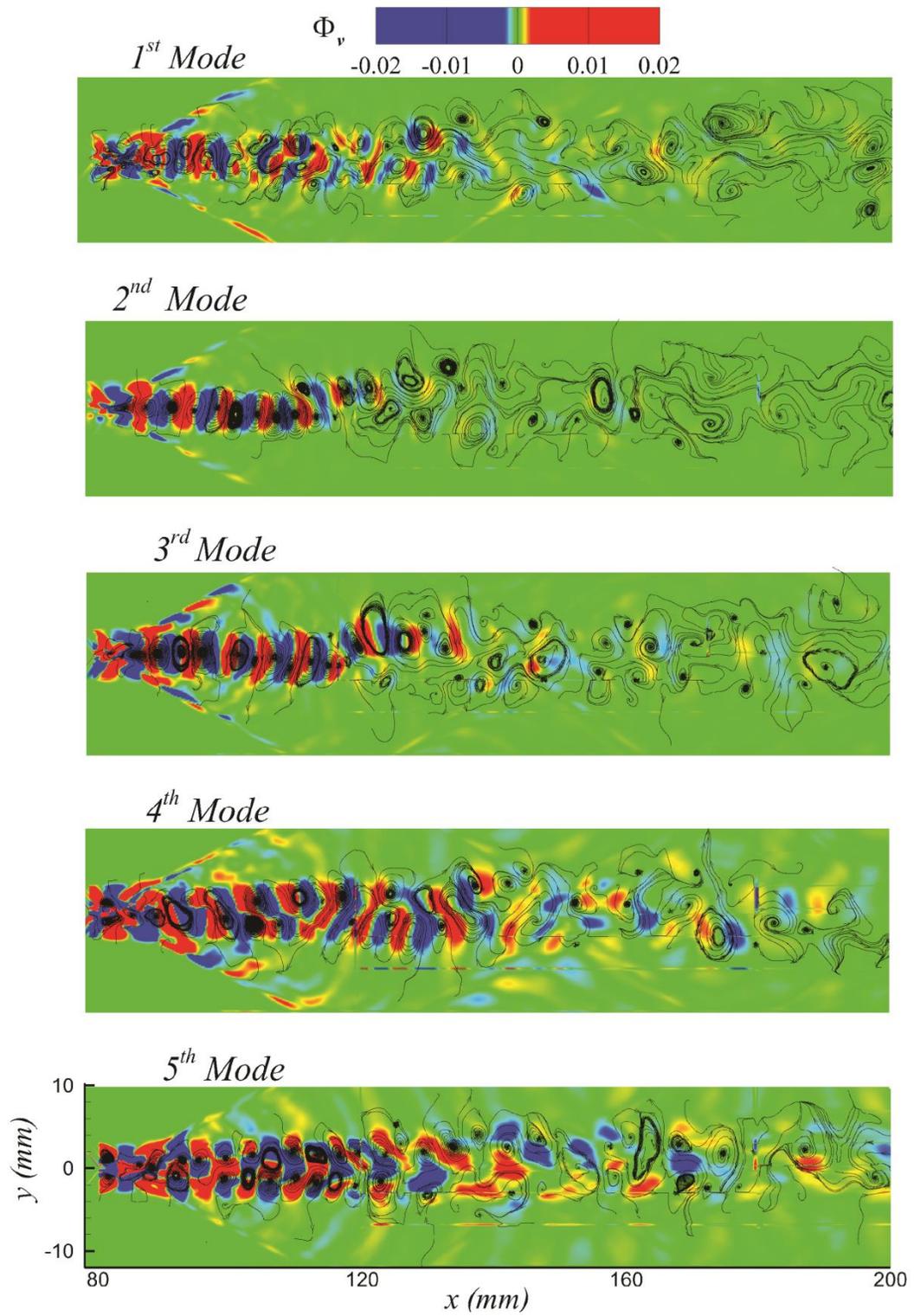

FIG. 14. First five most energetic modes of $SS-0.6$ case for energy based decomposition, contour represents the component corresponding transverse velocity

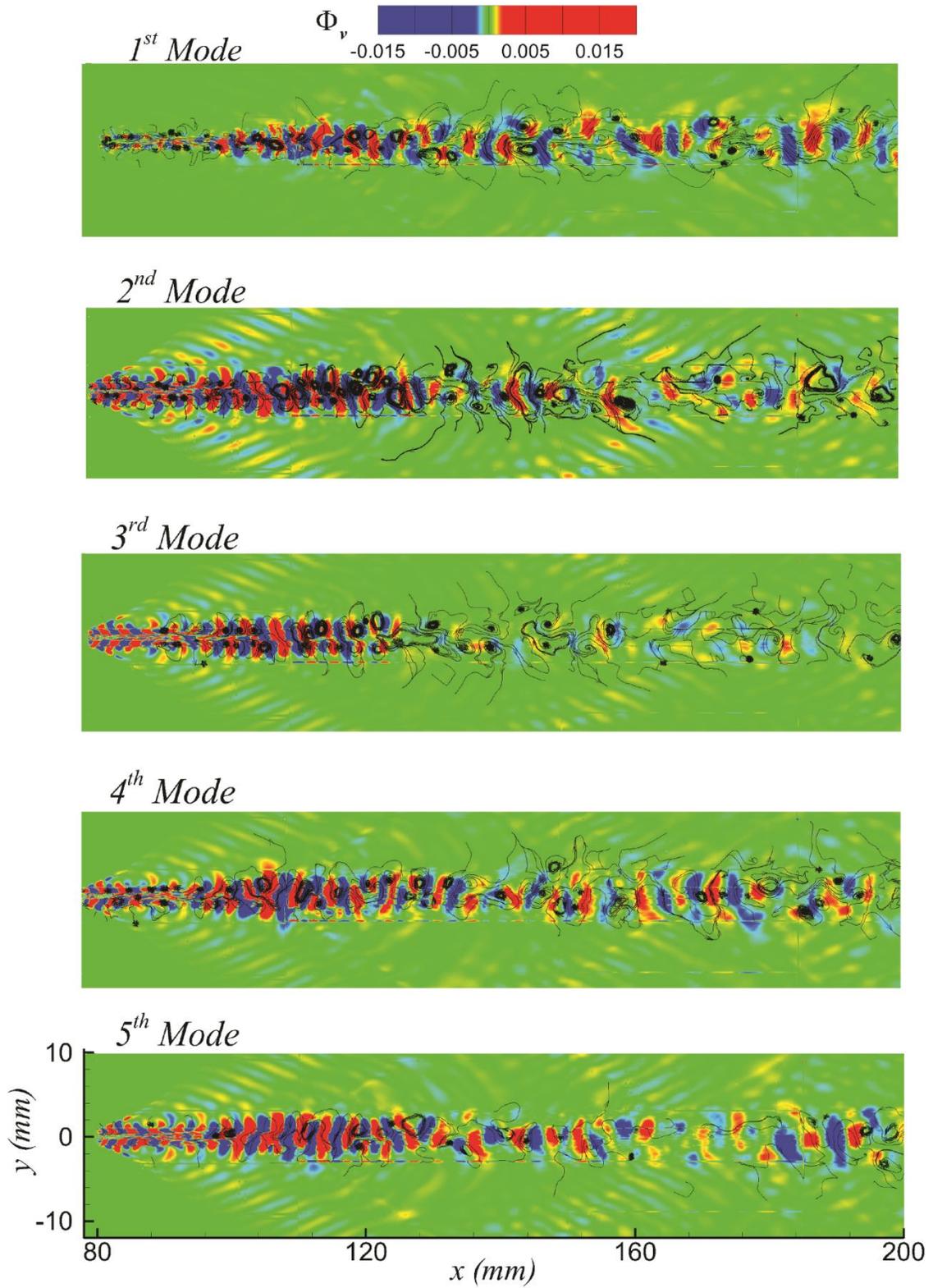

FIG. 15. First five most energetic modes of *TS−0.6* case for energy based decomposition; contour represents the component corresponding transverse velocity

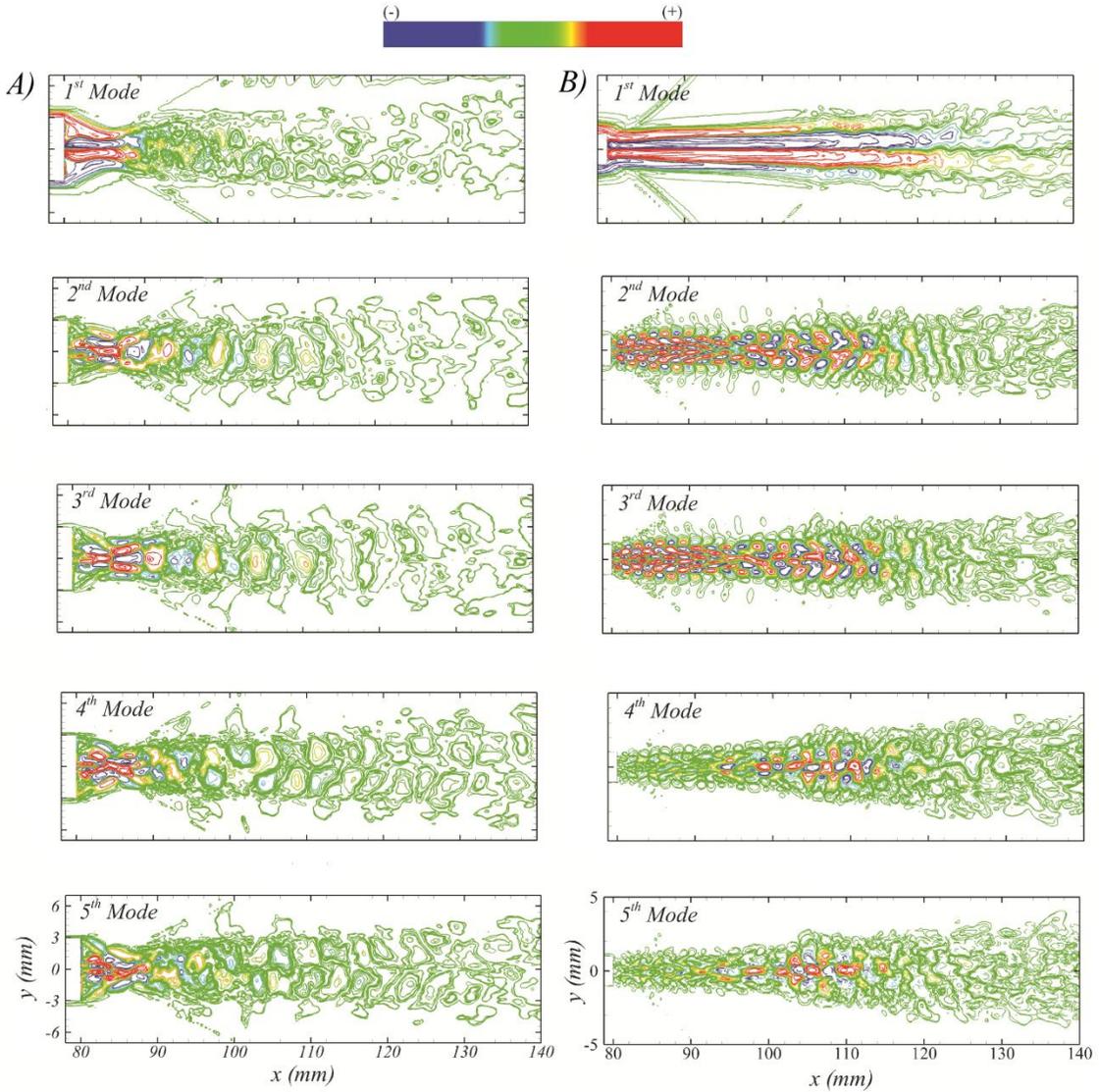

FIG. 16. Most energetic modes of vorticity vector based decomposition for A) *SS-0.6* & B) *TS-0.6* case

In Fig. 15 first five Eigen modes for the *TS−0.6* case is presented, the first mode represents the 99% energy distribution and hence depicts the mean motion. The presence of *K−H* vortices is clearly seen in the near field especially before the crossing shock wave system. The jet starts to become unstable around *X = 105 mm*, from the instantaneous vorticity contour plot it is evident that the amplification of the vortical structures starts at the crossing wave system. After this point the vortex breakdown becomes visible and the effect of delayed breakdown manifests itself into the reduced mixing efficiency and spreading. Once again the modes from 2-5 although carrying low energy content shed light into the shear layer flapping, the largest structures are mostly present throughout the modes along with small-scale structures. For both the cases, past breakdown point the behavior is very much similar to wake motion mimicking the Karman vortex street. For TS-0.6 case, the largest structures present in the flow vary between *8D_j−11D_j* and average large-scale structures range is comparable to the SS−0.6 cases.

The major difference observed in the jet spreading is mainly due to the lip thickness, since the increase in lip thickness results in increased shear layer thickness but at the cost of increased total pressure loss due to the recompression shock[4,5]. It is already known that the lip thickness in the base region of strut has a strong influence on the shape and size of the large-scale structures. Also the coherence and size of dominating structure increase with the lip thickness. This explanation also holds well in the present scenario and that's why the differences in the flow structures are observed for *TS & SS* cases.

Apart from the energy (velocity) based POD, enstrophy based decomposition is also performed for both (*TS & SS*) cases and reported in Fig. 16. Once again first 5 Eigen modes are reported for both cases since the $1^{st}$ mode has 100% energy distribution as it mimics the mean motion. From Fig. 16 it is apparent that $2^{nd}$ and $3^{rd}$ Eigen modes appear in pair with a spatial shift, the presence of various vortical structures on the either side of the shear layer is evident. The structures of the outer shear layer are *K−H* vortices while those embedded inside are due to the shearing motion between recirculation regions and shear layer. The *K−H* vortices appear to preserve the scale further downstream, however, those embedded ones start stretching earlier. The $2^{nd}$ – $3^{rd}$ modes have antisymmetric Eigen modes distribution along the centerline which suggests the symmetric small-scale spatial structures whereas $4^{th}$ & $5^{th}$ Eigen modes have symmetric modes shape distribution. The symmetric mode distribution actually suggests the antisymmetric shedding of vortical structures and is representative of the wake mode dominated by the Karman vortex street.

Contrasting the Eigen modes of *SS−0.6* with that of *TS-0.6* the striking difference is unraveled: here $2^{nd}$ & $3^{rd}$ modes have symmetric distribution representing the wake motion with relatively larger structures downstream, whereas $4^{th}$ & $5^{th}$ modes suggest antisymmetric distribution in the near field. Apart from the Eigen modes distribution, it appears that there is a convection of larger and smaller vortical structures of an alternating sign along the shear layer (see Fig. 11). The neck region where shear layer converges before growing outward and crossing shock system is accurately resolved in the Eigen modes. Another interesting feature revealed by $4^{th}$ & $5^{th}$ Eigen modes is the aggressive flapping of the jet which results in vortex breakdown which can also be verified from Fig. 11(A). This suggests that POD does not merely reveal the spatial coherence in turbulent flow field but is also representative of the series of events. Kostas[43] also suggested that Eigen modes may not necessarily represent the coherent structures only but also unfurl the events related to the fluctuating velocity & vorticity fields.

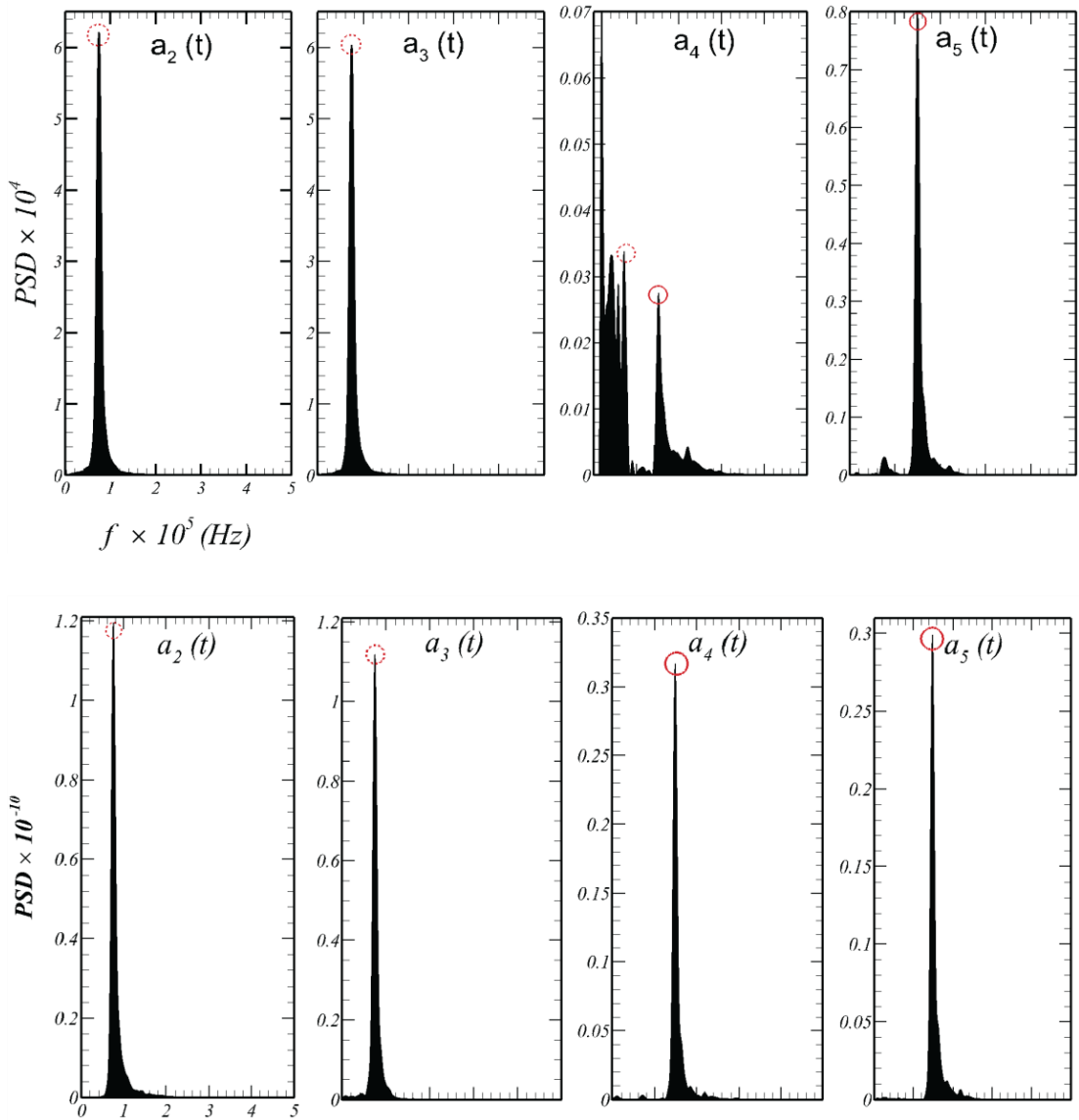

FIG 17. Spectra of POD temporal coefficients corresponding to energy (top) and enstrophy (bottom) for *SS-0.6* case at $M_c$-*1.4* (Solid circle: 147 KHz and Dashed circle: 74 kHz)

To quantify the same, the spectral analysis of the temporal coefficients for *SS-0.6* ($M_c$-1.4) case presented in Fig. 17 demonstrates the presence of at least two dominating frequencies (highlighted as a circle, refer figure caption for details). The temporal coefficients obtained for both energy and enstrophy based POD are utilized to perform the fast Fourier transform. Power spectra for 2[nd] & 3[rd] modes for both the approaches reveal the similar frequency of approximately 74 KHz whereas 4[th] & 5[th] again being same correspond to 147 KHz. This brings out the fact that 4[th] & 5[th] mode may be the super harmonic of the 2[nd] & 3[rd] modes. The 4[th] mode for energy based POD is dominated by various frequencies which is due to the fact that POD modes being orthogonal in a spatial sense are representative of the mixed frequencies. This suggests that the modes are strongly correlated at these two frequencies with the lowest frequency, i.e. 74 KHz being fundamental. The fundamental frequency is associated with the periodic flapping of shear layer due to the advection of structures mostly of *K-H* type whereas the first harmonic is attributed to the vortical shedding past the vortex breakdown point which mimics von Karman vortex street.

**D. Dynamic Mode Decomposition**

As mentioned and noted earlier, the POD decomposition can provide the spatial coherence of the vertical structures, but it does not reveal any information pertaining to the temporal variation. Therefore, in this section, the results of dynamic mode decomposition is presented and discussed for *TS − 0.6 & SS − 0.6* ($M_c$ - *1.4*) cases. The two most crucial parameters for the successful decomposition are temporal separation ($\Delta t$) between the snapshots and length of the snapshot sequence, i.e. spatial resolution. Hence, it is quite natural for extraction of pertinent flow process; the sampling frequency ($F_s$) must be sufficiently high. However, it is very much essential to comply with the Nyquist criterion which suggests that only those processes can be identified which are sampled with at least twice their inherent frequency. In the present study, the data is sampled at 1000 kHz, the choice of this sampling frequency comes from the FFT of the fluctuating velocity field. The vorticity magnitude field is utilized for the modal decomposition for both the cases as the choice of the variable have least or negligible effect on the output.

To demonstrate the convergence of the dynamic modes, decomposition is performed with four sets of snapshots, i.e. *N =* 40, 80, 100 & 121 on the SS−0.6 case. For all the given *N*, companion matrix (*S*) has *(N − 1) × (N − 1)* dimensions also the spatial extent is maintained same for all the cases. The Eigen spectrums for different *N* are mapped onto the familiar complex half-plane, and $\Delta t$ is the temporal separation between two consecutive snapshots. Upon evaluating the convergence of spectra, the detailed analysis is reported for the highest values of *N*, i.e. 121.

In Fig. 18 $L_2$ norm is presented, the symmetry of plot about the frequency axis is due to the processing of real-valued data, complex-valued data will result in asymmetric spectra about the imaginary axis. The mode corresponding to $0^{th}$ frequency is the representation of mean flow field, five distinct peaks are apparent from Fig. 18. The peaks possessing dominant frequency are highlighted with the yellow circle, where the first peak represents the fundamental frequency ($f_0$) and higher frequencies are the super harmonics. The first two frequencies are also witnessed from the Fourier transform of the POD temporal coefficients (Fig. 17). Here only modes corresponding to first two frequency, i.e. fundamental ($f_0$ =74 KHz) and first harmonic ($f_1$=148 KHz) is presented as the higher frequencies are the representative of the small-scale motion.

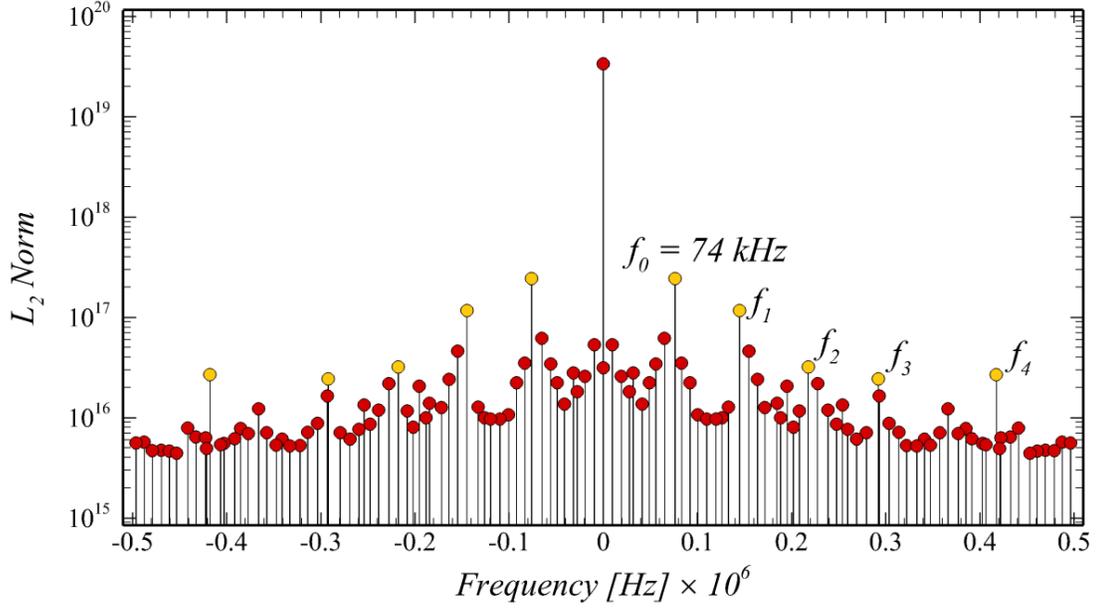

FIG. 18. $L_2$ norm of dynamic modes and corresponding frequencies

The dominant modes corresponding to fundamental and first harmonic is displayed in Fig. 19 for both the cases with real and imaginary parts presented separately using the contour plot. Here $\zeta_0$ & $\zeta_1$ correspond to $f_0$ and $f_1$, respectively in Fig. 19. For both modes, it is found that there is 90° phase shift which is consistent with the observation of the vorticity POD (Fig. 16). The mode corresponding to $f_0$ clearly suggests the presence of the spanwise rollers in the near field accompanied by the vortical tilting downstream, while the presence of structures along the shock is also revealed. This mode appears analogous to the 2nd & 3rd POD modes (Fig. 16). If the contour plot is followed carefully it appears that initially, vortical structures start to tilt at approximately two oscillatory periods from the reattachment shock. Beyond this point at around five oscillatory periods from the neck region of jet the vortices are stretched and appear separated. It can be inferred that this mode being fundamental is mostly responsible for advection-diffusion of large-scale structures which are also responsible for the exchange of the mass across the shear layer, in turn, the mixing efficiency (Fig. 6). Further downstream, the effect of vortical stretching becomes more evident with the presence of small-scale vortices along the edge of the shear layer.

The mode corresponding to $f_1$ differs significantly from the previous observation in a sense that the near-field distribution is mostly symmetric about the jet axis with asymmetry in the far field. This mode can be directly compared with the 4th & 5th vorticity based POD modes (Fig. 16). In the near-field region, it appears that the vortices are elongated in the normal direction and this could be due to the merging and pairing process of spanwise rollers which results in large-scale structures in the near field. Presence of small-scale vortices located along the edge of the jet boundary is witnessed only in the near-field which could be the $K-H$ vortices fed into the base region after the boundary layer separation. At around $X = 90$ mm, i.e. approximately one and a half oscillatory period from the neck, the merging of these smaller vortices with larger structures becomes evident. Due to the merging/pairing process, the two shear layer formed at the base region cease to exist as a single shear region at the downstream. So it seems that $\zeta_0$ is mostly representative of dynamics governing the stretching and tilting of large-scale vortices along with the presence of recirculating structure in the base region. More specifically, the antisymmetric spatial distribution suggests the vortical shedding representing Von Karman street vortices. However, $\zeta_1$ represents the super harmonic of the $\zeta_0$ with merging, pairing and eventually breakdown of the structures. The dominance of relatively more small structures in $\zeta_1$ is more pronounced but overall it represents a wide range of length scales.

Apart from the structures in jet and shear region, wave-like pattern is visible in the outer region, similar to Mach wave emanating from the jet boundary. In $\zeta_0$ the wavelength of the structures is large compared to the $\zeta_1$, some of these structures arise due to the interaction of crossing shock wave. The mode $\zeta_1$ is associated with the inner shear layer formed between the jet and recirculation region, where the instability in the near field is observed due to the addition of vortical structures formed at the base, which is basically $K-H$ type vortices convected from the base and perturb the shear layer to introduce the flapping motion. The convergence of shear layer towards the jet centerline is visible and post the reattachment region shear layer expands in the transverse direction, while at the further downstream merger of both shear layers due to the pairing of vortices is also observed.

Moving on for the *TS-0.6* case, the DMD results are also reported in this section, where Figs. 20 present the $L_2$ norm. Fig. 20 reveals that only two dominant frequencies are present, unlike the SS−0.6 case where super harmonics noticed. The two dominant modes corresponding to $f_1$ = 7.32 kHz and $f_2$ = 202 kHz are highlighted in the yellow color. It can be inferred that the first frequency is mostly responsible for the overall oscillatory motion experienced by the jet in a composite manner, while the $f_2$ may be responsible for wake motion. Once again the real and imaginary part of the dynamic modes corresponding to the highlighted frequencies is presented in the Fig. 21.

The $\zeta_1$ mode corresponding to $f_1$ does not reveal much information related to structure dominance in the flow field; however, it appears that this mode is associated with the flapping of the shear layer. In the near-field region, jet potential core and shear layer are clearly visible along the edge of the outer shear layer where the high-frequency structures become visible. Further downstream at around $X$ = 110 mm, some low-frequency structures are observed which are also present further downstream. It is possible that the jet breakup at this point leads to the formation of some large-scale structures due to vortices undergoing tremendous stretching. Whereas $\zeta_2$ corresponding to the higher frequency in the DMD spectrum represents the small-scale motion. The representation of this mode can be directly compared to the results of the POD Eigen modes.

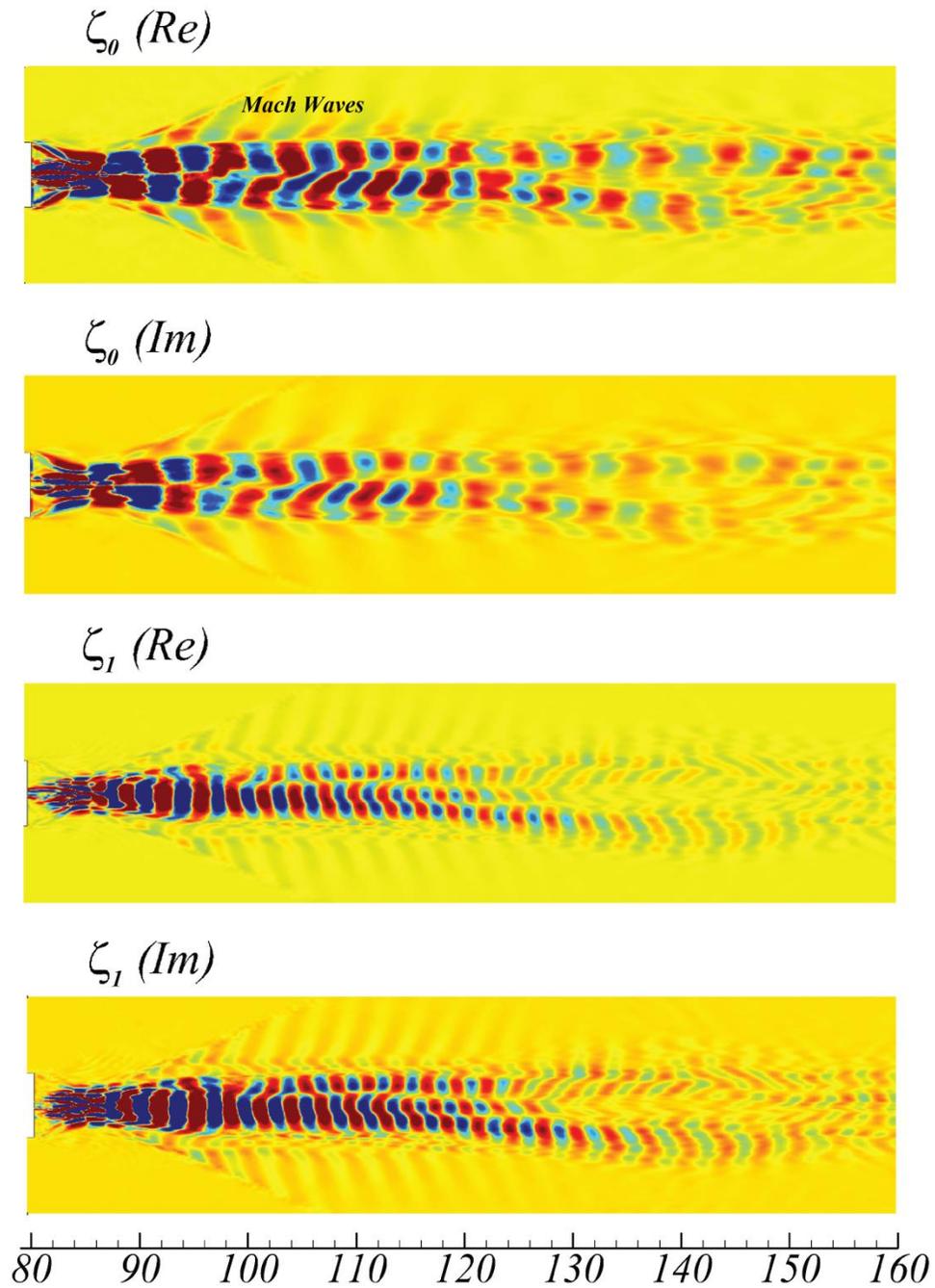

FIG. 19. Imaginary and real part of the dynamic modes corresponding to fundamental frequency ($f_0$) and first harmonic ($f_1$) for *SS-0.6* at $M_c$-1.4

The presence of *K−H* vortices along the shear layer can be clearly seen, further, this mode reveals the presence of structures along the inner region of the shear layer while pairing and merging of vortices becomes visible at around $X = 100$mm. Beyond this region, the oscillation of jet amplifies due to amplification of the instabilities introduced by the *K−H* type vortices present on the outer edge of the shear layer, and the same can be verified from the Fig. 11. At around $X = 110$ mm, vortex breakdown initiates and wake type flow is experienced with periodic shedding. It can be concluded that $\zeta_2$ is the most important mode representing most of the dynamics.

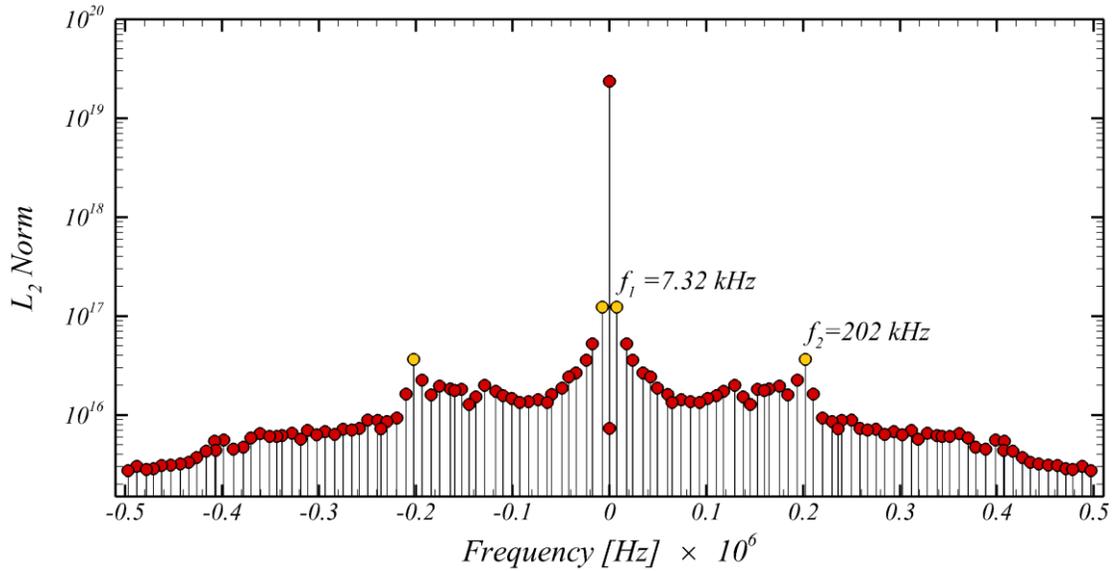

FIG. 20. $L_2$ Norm of dynamic modes against the frequencies for $TS - 0.6$ case, dominating modes highlighted in yellow

The instantaneous analysis combined with the modal decomposition techniques shed light into the various underlying physics that control the mixing of two streams in this type of configuration. There is enough evidence which suggests that convective Mach number alone is not sufficient for the characterization of the flow field, especially in the co-flow type situation. Whereas, velocity gradient is found to be an important parameter that affects the overall development of the large-scale structures which aid in mixing enhancement. Also, the higher gradient leads to increased turbulence along the shear layer which reflects in the near field mixing behavior.

Overall, the higher gradients at the base are observed across the shear layer due to the lip thickness and that essentially dominate the flow behavior. This leads to the conclusion that $M_c$-1.4 cases perform better when compared to $M_c$-0.37. The unsteadiness in the recirculating region introduces periodic shedding of the recirculation bubble which introduces the vigorous jet flapping leading to the breakdown of the jet. One very important observation from this exercise is the early breakdown of jet vortices, i.e. the near-field breakdown promotes mixing due to the formation of large-scale structures.

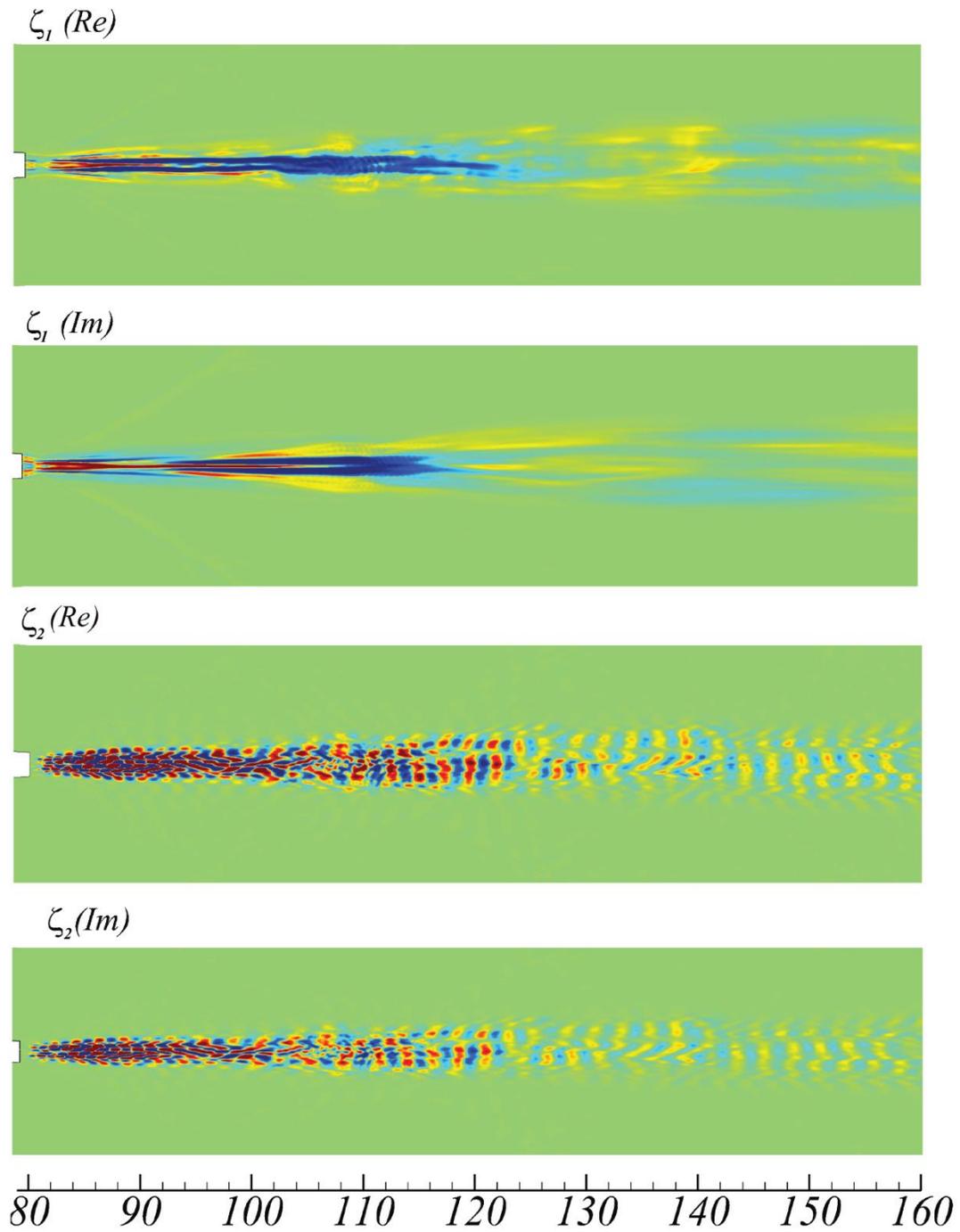

FIG. 21. Imaginary and real part of the dynamic modes corresponding to fundamental frequency ($f_1$) and first harmonic ($f_2$) for *TS-0.6* at *$M_c$-1.4*

## IV. CONCLUSION

Large eddy simulation for straight and tapered strut geometry with rectangular jet injection at two $M_c$ - *0.37* & 1.4 with 0.6 & 1 mm jet height is investigated in a Mach 2 co-flowing air. The mean data and grid independence results follow experiments closely, where the grid resolution is verified through LES quality index. The plots of the density gradient, vorticity, and Q-criterion depict the presence of various large-scale structures along the shear layer along with the pairing and merging of the vortical structures. It is found the effect of lip height in the base region and the velocity ratio play a critical role in the mixing layer growth and efficiency. As opposed to the general notion of convective Mach number concept the velocity gradient is observed to be responsible for mixing in the present cases studied herein. Also, $M_c-1.4$ case performs better for both the geometrical variations compared to the $M_c -0.37$ case, where the difference in the spatial scales of the structures is clearly witnessed. To characterize the spatial dominance of the vortical structures, energy as well as the vorticity based POD is performed. Both the approaches reveal the underlying mechanism which is responsible for the breakdown and the mixing behavior. The Eigen modes shed light into the advection-diffusion of the vortices through different modes, mostly large-scale structures are found to be spread more in straight configuration than tapered strut. These large-scale structures pull the mass outside of the shear layer toward the jet core at the same time facilitating the expulsion of mass from jet core to outward region. The dynamic mode decomposition is also performed to separate the coherence on the basis of pure frequency and understand the governing dynamics of the flow. The DMD spectra reveal the multiple harmonics for the straight case whereas no such phenomenon is present for the tapered case. The results of DMD are also consistent with the observation of the POD modes. Overall both the modal decompositions offer aid in understanding the breakdown, pairing/merging and stretching of vortices which directly affect the mixing characteristics for a given configuration studied herein.

**ACKNOWLEDGEMENTS**

Financial support for this research is provided through IITK-Space Technology Cell (STC). Also, the authors would like to acknowledge the High-Performance Computing (HPC) Facility at IIT Kanpur ([www.iitk.ac.in/cc](www.iitk.ac.in/cc)).